\documentclass[prd,twocolumn,showpacs,nofootinbib,amsmath,amssymb,floatfix,superscriptaddress,showkeys]{revtex4-1}
\usepackage{multirow}
\usepackage{graphicx}
\usepackage{epstopdf}
\usepackage{dcolumn}
\usepackage{bm}
\usepackage{threeparttable}
\usepackage{subfigure}
\usepackage{color,txfonts}
\usepackage{ulem}
\definecolor{blue0}{rgb}{0,0,0.6}
\usepackage[colorlinks,linkcolor=blue0,anchorcolor=blue0,citecolor=blue0,urlcolor=blue0]{hyperref}

\newcommand{\jcap}{J. Cosmol. Astropart. Phys.}
\newcommand{\physrep}{Phys. Rep.}
\newcommand{\mnras}{Mon. Not. R. Astron. Soc.}
\newcommand{\araa}{Annu. Rev. Astron. Astrophys.}
\newcommand{\aap}{Astron. Astrophys}

\newcommand{\apjs}{Astrophys. J. Suppl.}

\begin{document}

\title{Revisiting the search for dark matter subhalos using the Fermi-LAT 4FGL-DR4 catalog}
\author{Ji-Gui Cheng}
\email{cheng-jg@hnust.edu.cn}
\affiliation{School of Physics and Electronics, Hunan University of Science and Technology, Xiangtan 411201, China}
\affiliation{Hunan Provincial Key Laboratory of Intelligent Sensors and Advanced Sensor Materials, Hunan University of Science and Technology, Xiangtan 411201, China}
\author{Le Zou}
\affiliation{Department of Physics, Xiangtan University, Xiangtan 411105, China}
\affiliation{Key Laboratory of Stars and Interstellar Medium, Xiangtan University, Xiangtan 411105, China}

\date{\today}

\begin{abstract}
Numerical simulations suggest that dark matter halos surrounding galaxies host numerous small subhalos, which might be detectable by the Fermi-LAT. In this work, we revisit the search for gamma-ray subhalo candidates using the latest Fermi-LAT 4FGL-DR4 catalog. The search is performed by fitting the spectral data of unassociated point sources in the catalog through an unbinned maximum likelihood method. We consider two models in the fitting. One is an empirical function provided by the catalog, and another is a DM model in which DM particles within nearby subhalos annihilate into gamma rays and other Standard Model particles. Based on the fitting results, we identify 32 candidates for which the maximum likelihood value of the DM model fit exceeds that of the empirical function fit. The estimated J-factors of these candidates range from $0.2$ to $5.8 \times 10^{20}\,{\rm GeV^{2}\,cm^{-5}}$, the DM particle masses vary from $30$ to $500\,{\rm GeV}$ and 12 of them are within the range of $[30, 80]\,{\rm GeV}$. Candidate 4FGL J2124.2+1531 is an exception with a J-factor of $4.52 \times 10^{21}\,{\rm GeV^{2}\,cm^{-5}}$ and a particle mass of $3108.44\,{\rm GeV}$. Interestingly,  the identified candidates do not overlap with those reported in previous works, and we discuss the possible reasons for the discrepancy. At the current stage, we cannot rule out the possibility that these candidates are gamma-ray pulsars, and further confirmation through multi-band observations is required.
\end{abstract}

\maketitle

\section{Introduction}
\label{sec:introduction}
Dark matter (DM) is one of the most intriguing components of the Universe, with its existence strongly supported by various observational evidence, including galactic rotation curves, gravitational lensing, and cosmic microwave background (CMB) anisotropies \cite{1996PhR...267..195J,2005PhR...405..279B}. Within the prevailing cold dark matter cosmology model that successfully explains the CMB observations, the predicted density of DM is approximately 6.4 times that of baryonic matter \cite{2020A&A...641A...6P}. Despite this, the nature of DM remains unknown. Various theoretical candidates have been proposed, including weakly interacting massive particles (WIMPs), axion-like particles (ALPs), and primordial black holes (PBHs) \cite{2017FrPhy..12l1201Y}. Identifying these candidates and verifying their existence is a pressing challenge in DM research.

For many years, WIMPs have been a highly competitive candidate. Their mass typically falls in the ${\rm GeV}$ range. Studies suggest that these particles may produce detectable gamma-ray emissions through self-annihilation or decay, observable by telescopes like the Fermi-LAT \cite{2009ApJ...697.1071A,2010ARA&A..48..495F,2012ApJ...747..121A,2012PhRvD..86h3511A,2012JCAP...07..054B,2015JCAP...12..035B,2015PhRvD..91l2002A,2016PhRvD..93j3525L,2018PhRvL.120t1101A,2019JCAP...11..045C,2022SciBu..67..679D,2024PhRvD.109b3532D}. If WIMPs are thermal relics, the expected cross-section for s-wave annihilation is $\left \langle \sigma v \right \rangle \sim 3.0 \times 10^{-26}\;{\rm cm^3\;s^{-1}}$. The extremely weak interactions make WIMPs challenging to detect. Many works have been conducted to search for them, but no detections have been reported yet \cite{2015PhRvL.115w1301A,2017IJMPA..3230006K,2018PhRvL.121k1302A,2023PhRvL.131d1003A}. Consequently, strong constraints on the WIMP parameters have been established (e.g. \cite{2023PhRvD.108f3015C}). With prolonged searches that yield no detections and new signals emerging from experimental data (e.g. \cite{2020PhRvD.102g2004A}), ALPs and other DM candidates are attracting increasing attention from the community. However, WIMPs remain a compelling target for gamma-ray searches.

Numerical simulations on the formation of large structures in the Universe suggest that DM halos on the scale of galaxies host numerous small subhalos \cite{2008MNRAS.391.1685S}. These subhalos are promising sites for detecting WIMP annihilation in the gamma-ray band, as they are too small to capture significant amounts of ordinary matter and are invisible to other electromagnetic bands. The detectability of these subhalos depends on various factors, including the abundance of subhalos within the Milky Way, the density profile of the subhalos and the host halo, and the sensitivity threshold of the observational instruments. Many theoretical works have been conducted using simulations or semi-analytical methods to assess the detectability of the Fermi-LAT and future instruments like the CTA \cite{2016JCAP...09..047H,2019Galax...7...60H,2022PhRvD.106h3023F}. These studies suggest that only a few subhalos would likely be bright enough for detection, even though they are expected to be abundant in our galaxy.

The Fermi-LAT source catalog offers a valuable dataset for searching for potential DM subhalos. Ref.~\cite{2015JCAP...12..035B} performs a systematic analysis using the third Fermi-LAT catalog (3FGL, \cite{2015ApJS..218...23A}) to search gamma-ray subhalo candidates. This study identifies 24 high-latitude candidates from the unassociated 3FGL point sources whose photon spectra align with expectations for WIMPs annihilation via the $b \bar{b}$ channel, with inferred DM mass ranging from $\sim 20$ to $70\;{\rm GeV}$. More recently, machine learning techniques have been employed in the searches. Ref.~\cite{2023MNRAS.520.1348G} suggests that no DM subhalo candidates can be identified within the fourth Fermi-LAT catalog (4FGL, \cite{2020ApJS..247...33A}). Ref.~\cite{2023JCAP...07..033B}, however, estimate that some candidates may be present in the 4FGL-DR3 catalog (the third data release of the 4FGL). The expected number of candidates depends on the DM mass and a threshold value related to the label of Bayesian neural network classification.

The Fermi-LAT catalog has now been updated to the fourth data release, 4FGL-DR4, which includes 14 years of accumulated data \cite{2023arXiv230712546B}. This latest release contains 7194 gamma-ray sources, including 2044 unassociated point sources (twice the number in the 3FGL catalog and 309 more than in the previous 4FGL-DR3). With this expanded dataset, it is compelling to explore how many DM subhalo candidates can be identified and to examine any variations in results compared to previous searches. In this paper, we are focusing on the unassociated point sources of the 4FGL-DR4 catalog, applying an unbinned maximum likelihood analysis to search for DM subhalos candidates. The paper is organized as follows. In Sec.~\ref{sec:model}, we show the calculation of gamma-ray spectrum produced by DM subhalos. In Sec.~\ref{sec:data}, we describe the details of the Fermi-LAT data analysis and the identification procedural of subhalo candidates. In Sec.~\ref{sec:rnd}, we report the list of identified candidates. The discussion and comparisons to previous studies are presented in Sec.~\ref{sec:dis}.

\section{Gamma-ray signals from DM subhalos}
\label{sec:model}
If a DM subhalo consists of WIMPs, the gamma-ray production $\phi_{\rm \gamma}(E_{\rm \gamma})\;({\rm cm^{-3}\;s^{-1}\;GeV^{-1}})$ from WIMPs self-annihilation can be estimated as
\begin{equation}
\label{eq:spectrum}
    \phi_{\rm \gamma}(E_{\rm \gamma}) = \frac{1}{4\pi} \frac{\left \langle \sigma v \right \rangle}{2 m_{\rm \chi}^2} \frac{{\rm d} N_{\rm \gamma}} {{\rm d} E_{\rm \gamma}} \times \mathcal{J},
\end{equation}
where $\left \langle \sigma v \right \rangle$ is the velocity-averaged cross section, $m_{\rm \chi}$ is the DM mass, and ${\rm d} N_{\rm \gamma} / {\rm d} E_{\rm \gamma}$ is the differential photon production per annihilation which can be obtained using PPP4DMID \cite{2011JCAP...03..051C}. While the J-factor (${\rm GeV^2\;cm^{-5}}$) can be calculated by line-of-sight integral along $l$ with respect the DM density profile $\rho(r)$, 
\begin{equation}
    \mathcal{J} = \int_{0}^{\Delta \Omega} \left\{ \int_{\rm l.o.s} \rho^2 \left[ r(d, l, \theta) \right] {\rm d}l \right\} {\rm d} \Omega.
\end{equation}
In this equation, $d$ is the distance between the earth and the subhalo center, $\theta$ is the angle between $l$ and $d$, and $r(d, l, \theta) = \sqrt{d^2 + l^2 -2dl \cos{\theta}}$ is the distance from the subhalo center. If $d$ is much larger than the spatial extension of the subhalo, the J-factor can be approximated as
\begin{equation}
    \mathcal{J} \approx \frac{\int \rho^2 {\rm d} V}{d^2}.
\end{equation}

For WIMPs with masses in the GeV range, Eq.~\ref{eq:spectrum} shows that the anticipated gamma-ray spectral shape of a subhalo depends on the choice of annihilation channel. The brightness of the gamma-ray emission is proportional to  the J-factor, and thereby strongly influenced by the density profile. Numerical simulations on DM halo formation suggest several profiles, including the famous Navarro-Frenk-White (NFW) profile \cite{1997ApJ...490..493N} and the Einasto profile \cite{2008MNRAS.391.1685S}. Some theories propose even denser profiles, such as the DM-spike \cite{2014PhRvL.113o1302F} and ultracompact minihalos \cite{2009PhRvL.103u1301S}. It can be expected that subhalos with denser profiles would be brighter since WIMPs annihilation would be more active in these regions.

\section{Identify DM subhalo candidates with the 4FGL-DR4 catalog}
\label{sec:data}
The identification process of DM subhalo candidates is based on two preliminary selection criteria. (1) The photon emission from WIMPs annihilation is expected to be detectable only in the gamma-ray band so that a DM subhalo should not have a counterpart in other wavelengths. (2) The gamma-ray flux from WIMPs annihilation within a DM subhalo should be steady, with no significant flux variations over time. We initially select sources from the 4FGL-DR4 catalog that satisfy both the criteria. Then, for each selected source, we perform an unbinned likelihood spectral analysis to assess whether its gamma-ray spectrum aligns with the expectations for a DM origin. The analysis details are described below.

\subsection{Data reduction}
\label{subsec:data_reduction}
The 4FGL-DR4 catalog provides comprehensive information on sources detected by the Fermi-LAT over 14 years of data accumulation, including their location, potential associations with sources observed by other instruments, flux variability, and spectral properties characterized by empirical functions like the power-law and log-parabola functions. Based on the catalog, we conducted a preliminary selection to exclude sources unlikely to be DM candidates. First, sources with associations were filtered out using the parameters {\tt ASSOC1} and {\tt ASSOC2}, and 2044 unassociated point sources remained. Next, we assessed flux variability using the {\tt Variability\_Index} parameter. We exclude 72 sources with ${\tt Variability\_Index} > 27.69$, indicating non-steady flux at a $99\%$ confidence level, following \cite{2023arXiv230712546B}. This left 1972 sources for further consideration ($1972/2044$). In the final step of the preliminary selection, we removed sources from the low Galactic latitude region (${ \left | b \right | < 20^{\circ}}$), as the background is complicated in this region and the Fermi-LAT sensitivity is lower. After this step, 644 sources remained ($644/1972$), warranting further spectral analysis.

Using the latest Fermi-LAT software {\it Fermitools 2.2.0}, we extract photon data for the 644 target sources adopting a relatively small region of interest (ROI) of $5^{\circ}$. That is to improve the efficiency of the analysis process. The time range of used data is from 2008 August 4 to 2022 August 2 (corresponding to MET time 239557417 to 681170000) to keep consistency with the 4FGL-DR4 catalog, with the photon energies cover from $100\;{\rm MeV}$ to $500\;{\rm GeV}$. Following the standard Fermi-LAT point source data analysis thread, we adopt the {\tt SOURCE} event class with ${\tt evclass} = 128$ and ${\tt evtype} = 3$. A maximum zenith angle of $z_{\rm max} = 90^{\circ}$ is applied to minimize contamination from the bright Earth limb, and a data quality cut ${\tt (DATA\_QUAL>0)\&\&(LAT\_CONFIC==1)}$ is employed to ensure the scientific reliability of the data. We use the default instrument response function, i.e. {\tt P8R3\_SOURCE\_V3}, to generate the exposure map for each ROI. The extracted data and the exposure maps will support the likelihood analysis described in the following section.

\subsection{Likelihood fitting}
\label{subsec:data_likelihood}
To further identify DM candidates from the 644 target sources, we perform spectral fitting for each source’s ROI using an unbinned likelihood method. The fitting is implemented using the {\it Fermitools} software, by which a photon-counts map is generated based on spectral models for all the sources inside the ROI, including the target point source, other resolved sources, and two unresolved background components: the Galactic diffuse emission and the extragalactic isotropic diffuse emission. Then, the theoretical photon-counts map is compared to the observational data to assess the appropriateness of the models and to derive the model parameters. During this process, the spectral models of each ROI are compiled into a single model file, which can be generated using the Python package {\it LATSourceModel} based on the 4FGL-DR4 catalog. Although we adopt a $5^{\circ}$ ROI, the model file includes sources within a $10^{\circ}$ radius. This is because photons from sources outside the ROI may also be detected, corresponding to the instrument’s point-spread function. 

The log-likelihood function used for the unbinned likelihood fitting can be expressed as 
\begin{equation}
	\begin{aligned}
		\ln &L(\theta_{\rm inc}, \theta_{\rm tar}) = \\
		& \sum_{\rm i=1}^{N} \ln [ F_{\rm tar}(E_{\rm i}, \theta_{\rm tar}) \epsilon_{\rm tar}(E_{\rm i}) + F_{\rm inc}(E_{\rm i}, \theta_{\rm inc}) \epsilon_{\rm inc}(E_{\rm i}) ] \\
		& - \int [ F_{\rm tar}(E, \theta_{\rm tar}) \epsilon_{\rm tar}(E) + F_{\rm inc}(E, \theta_{\rm inc}) \epsilon_{\rm inc}(E) ] {\rm d} E,
	\end{aligned}
	\label{eq:likelihood}
\end{equation}
where $F$ is the photon flux (${\rm cm^{-2}\;s^{-1}\;GeV^{-1}}$) predicted by the spectral model with parameter $\theta$, $E$ represents the photon energy, $N$ is the total photon counts in the ROI, $\epsilon$ is the exposure which depends on photon energy and source location, and the subscript ``${\rm tar}$" and ``${\rm inc}$" are corresponding to the target and incidental sources respectively. During the fitting procedure, since we use the same time range as in the 4FGL-DR4 catalog, the spatial positions of the sources are fixed according to the 4FGL-DR4 values. The initial parameters for the spectral models are also taken from the catalog. For resolved sources located beyond the $5^{\circ}$ radius of the ROI, we fix their spectral parameters to facilitate faster convergence of the target source parameters.

For the spectral model of the target source, we consider two distinct models: (1) an empirical function as provided in the 4FGL-DR4 catalog and (2) a DM annihilation model, which can be calculated using the {\tt DMfitFunction} compiled in the {\it Fermitools}. In the DM model, we consider the s-wave WIMPs self-annihilation so that the velocity-averaged cross-section is fixed at $\left \langle \sigma v \right \rangle = 3.0 \times 10^{-26}\;{\rm cm^{3}\;s^{-1}}$ for thermal relics, the $b \bar{b}$ annihilation channel is adopted to remain consistent with previous searches \cite{2012ApJ...747..121A,2015JCAP...12..035B}, and we do not separately model the diffuse emission from the Galactic DM halo since it is expected to be faint compared to the target source flux \cite{2022PhRvD.106h3023F}. For each of the two models, likelihood fitting is performed individually to derive the corresponding best-fit parameters and statistical values. The statistical values include the test-statistic (TS) value which indicates the confidence of the target source model, and the maximum log-likelihood value ($\ln L$). 

Our final selection criteria for identifying DM candidates are based on the statistical values: (1) the TS value for the DM model must be greater than 25 (${ \rm TS}_{\rm DM} > 25$), which corresponds to approximately a $5\sigma$ confidence level, suggesting the presence of a gamma-ray point source with a DM origin, and (2) the maximum log-likelihood value of the DM model must exceed that of the empirical function, i.e. ${\rm ln}L_{\rm DM} > {\rm ln}L_{\rm emp}$,  indicating that the data show a better consistency with the DM model. Applying these criteria, we identify 32 sources ($32/644$) as DM subhalo candidates.

\section{Results}
\label{sec:rnd}
\begin{figure*}
\centering
\includegraphics[width=1.5\columnwidth]{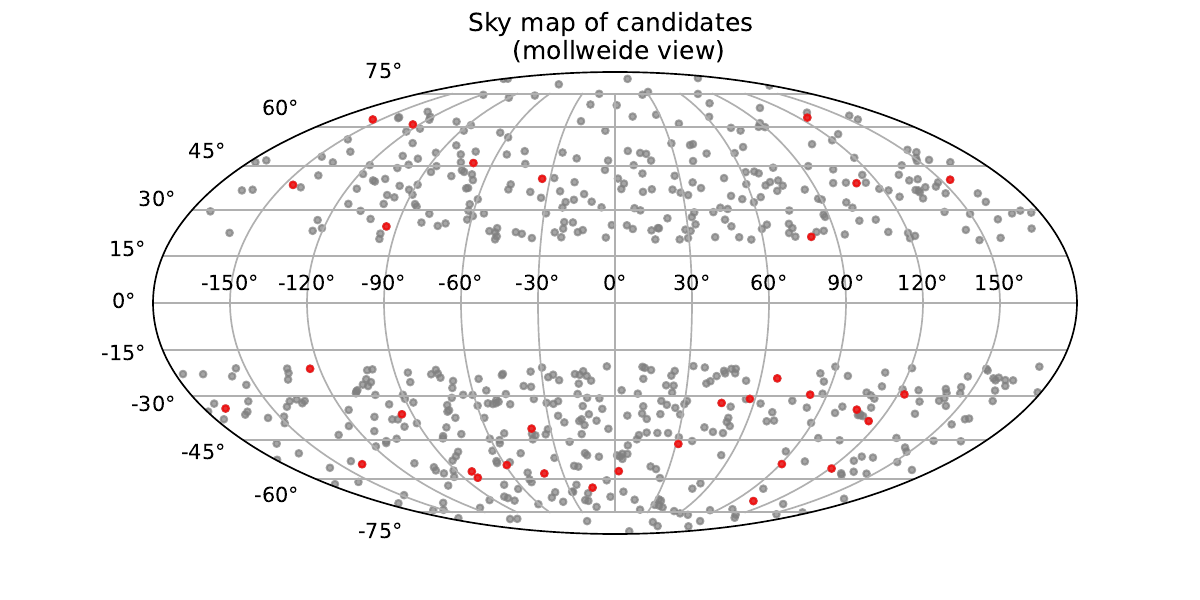}
\caption{The sky map of the 32 identified DM subhalo candidates marked with red dots {\it v.s.} 644 unassociated point sources from the 4FGL-DR4 catalog marked with gray dots.}
\label{fig:map}
\end{figure*}

We identify 32 high-latitude DM subhalo candidates from 2044 unassociated point sources in the 4FGL-DR4 catalog based on the results of the unbinned likelihood analysis. The distribution of these candidates in the sky is shown in Fig.~\ref{fig:map}, with 9 located in the northern sky and 23 in the southern sky. Detailed information about the sources and their best-fit parameters is provided in Tab.~\ref{tab:params}. As summarized in the table, the identified candidates are generally faint, with integral fluxes of $1$ to $100\;{\rm GeV}$ concentrated in the range $[0.7, 2] \times 10^{-10}\;{\rm cm^{-2}\;s^{-1}}$. The derived J-factors, which characterize the DM density distribution, span from $0.2$ to $5.8 \times 10^{20}\;{\rm GeV^{2}\;cm^{-5}}$. The best-fit DM masses for most candidates are below $500\;{\rm GeV}$, with 12 falling in the range $[30, 80]\;{\rm GeV}$, 14 in $[80, 300]\;{\rm GeV}$, and 5 in $[300, 500]\;{\rm GeV}$. One notable exception is candidate 4FGL J2124.2+1531, which exhibits an unusually high DM mass of $m_{\rm \chi} = 3108.44\;{\rm GeV}$ and a J-factor of $\mathcal{J} = 4.52 \times 10^{21}\;{\rm GeV^{2}\;cm^{-5}}$.

\begin{figure*}
\centering
\includegraphics[width=0.2\textwidth]{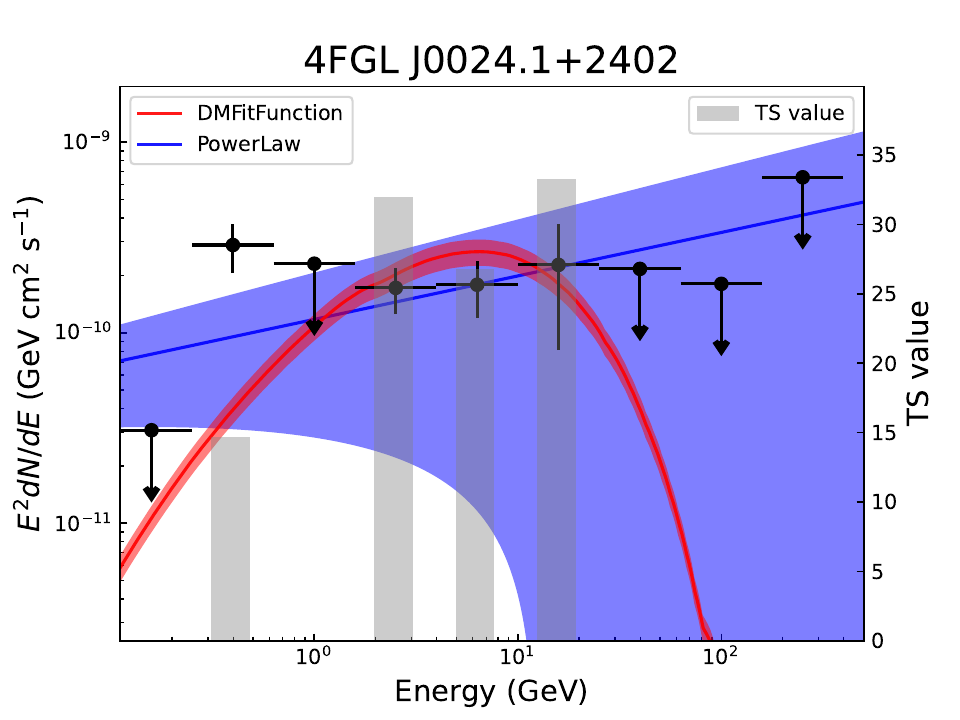}
\includegraphics[width=0.2\textwidth]{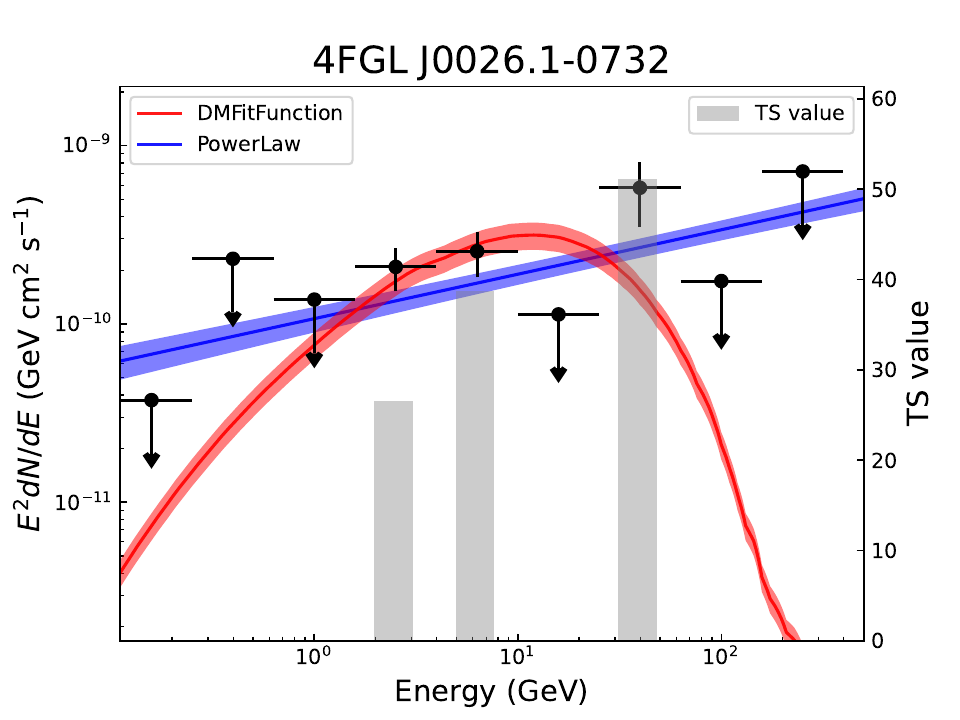}
\includegraphics[width=0.2\textwidth]{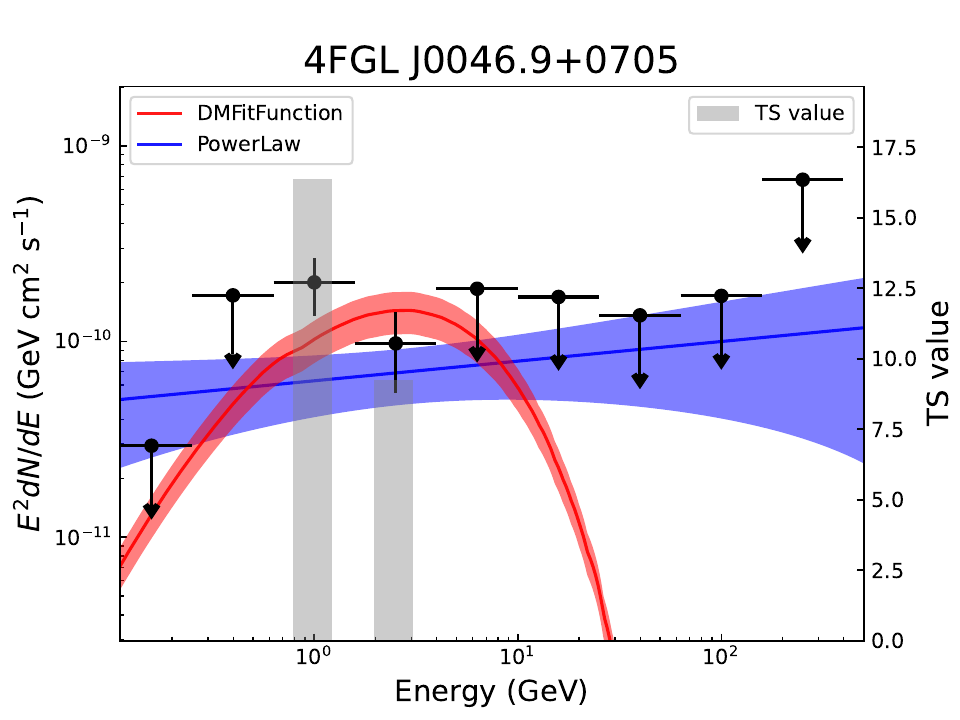}
\includegraphics[width=0.2\textwidth]{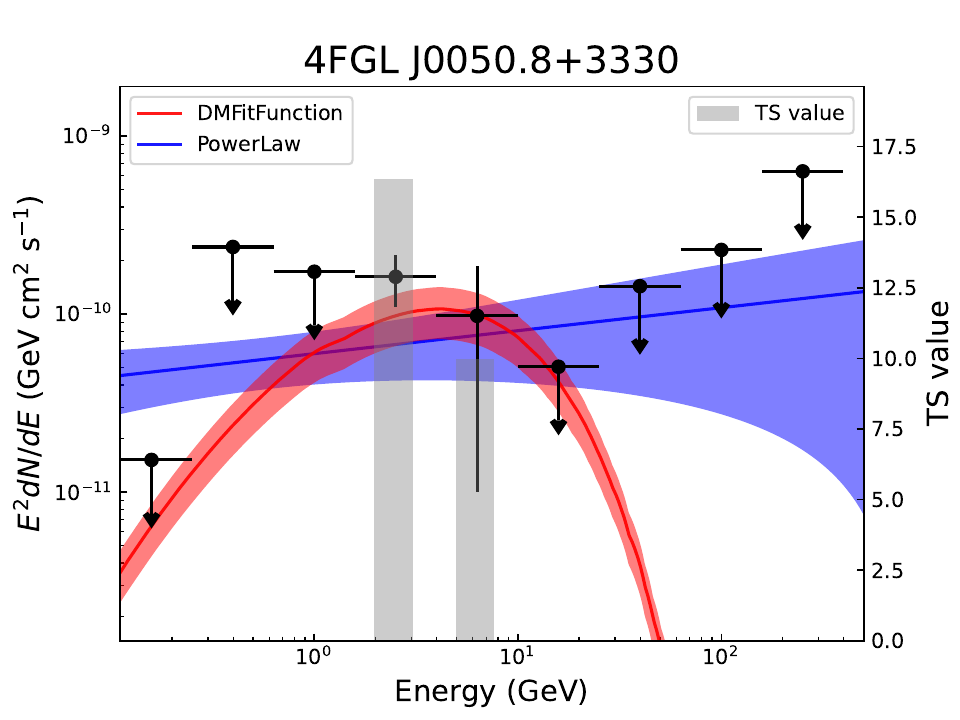}\\
\includegraphics[width=0.2\textwidth]{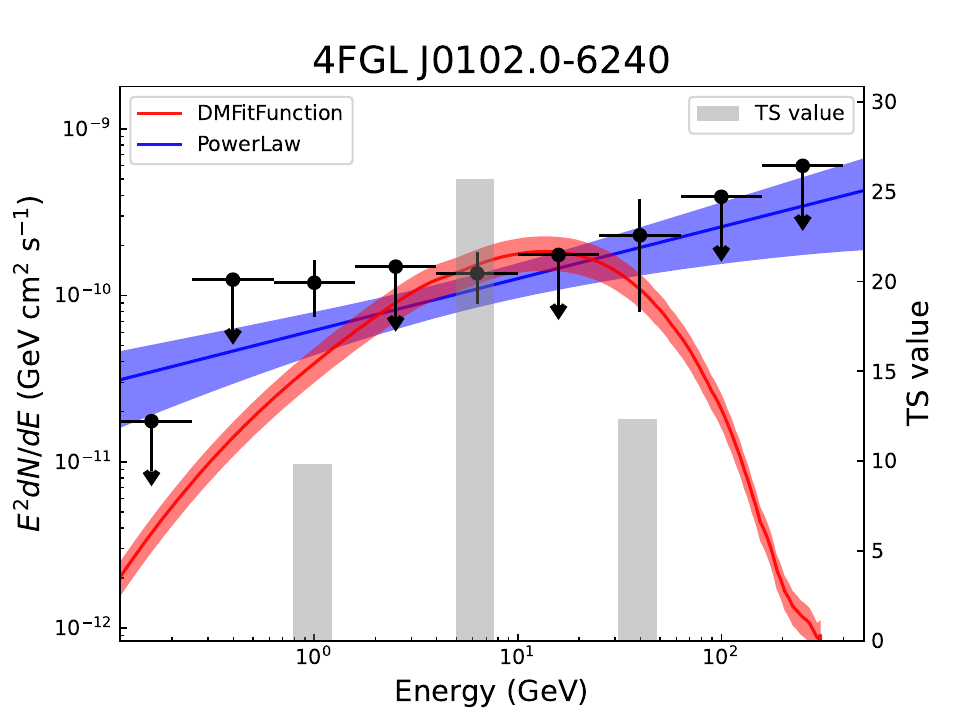}
\includegraphics[width=0.2\textwidth]{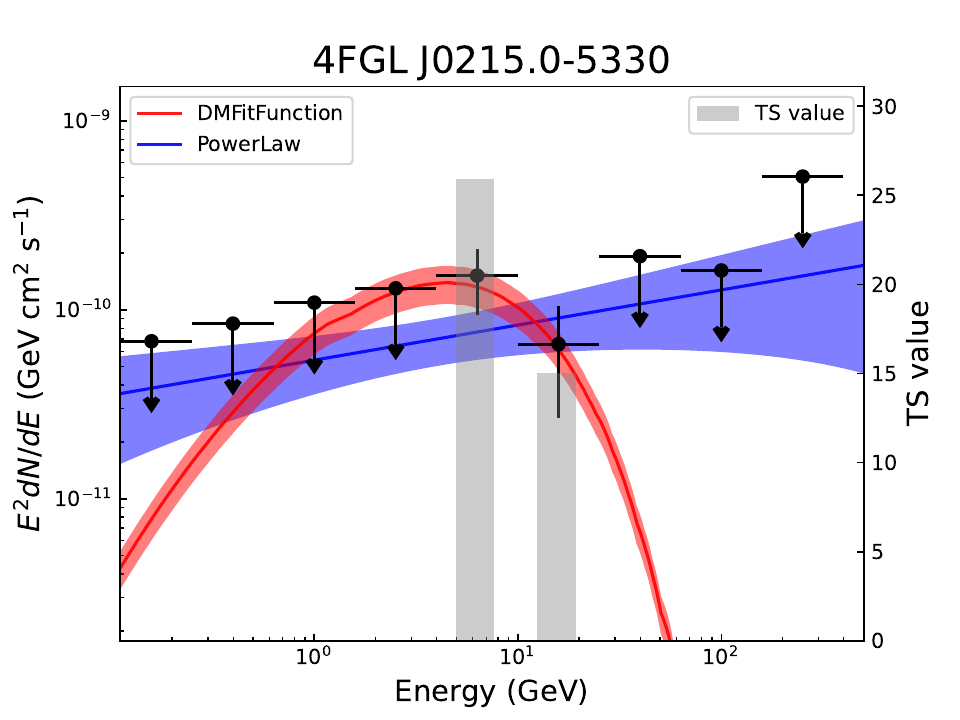}
\includegraphics[width=0.2\textwidth]{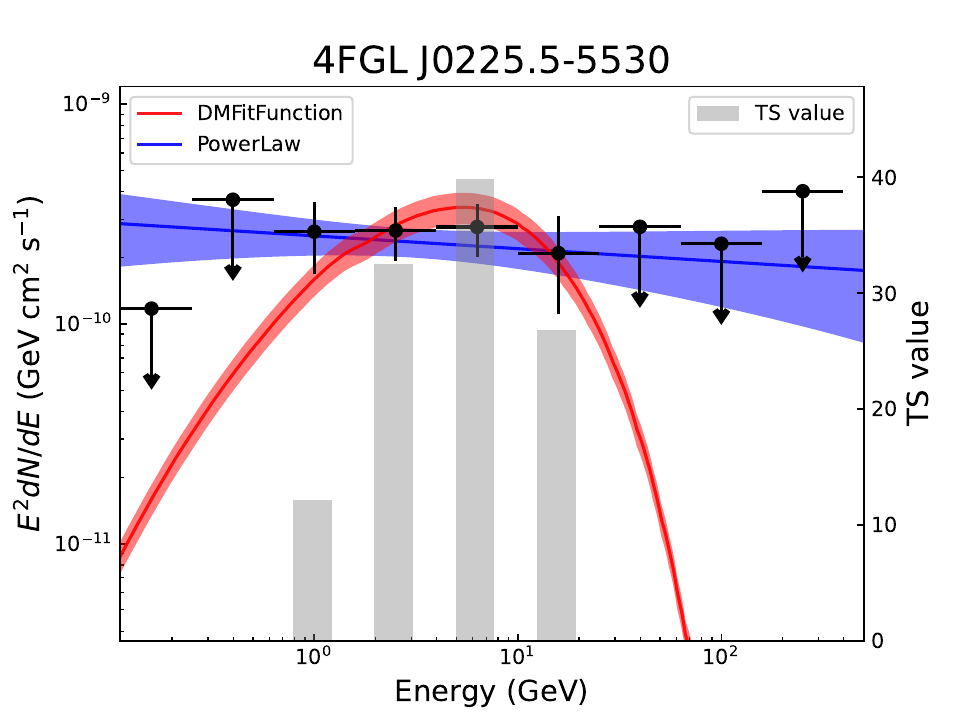}
\includegraphics[width=0.2\textwidth]{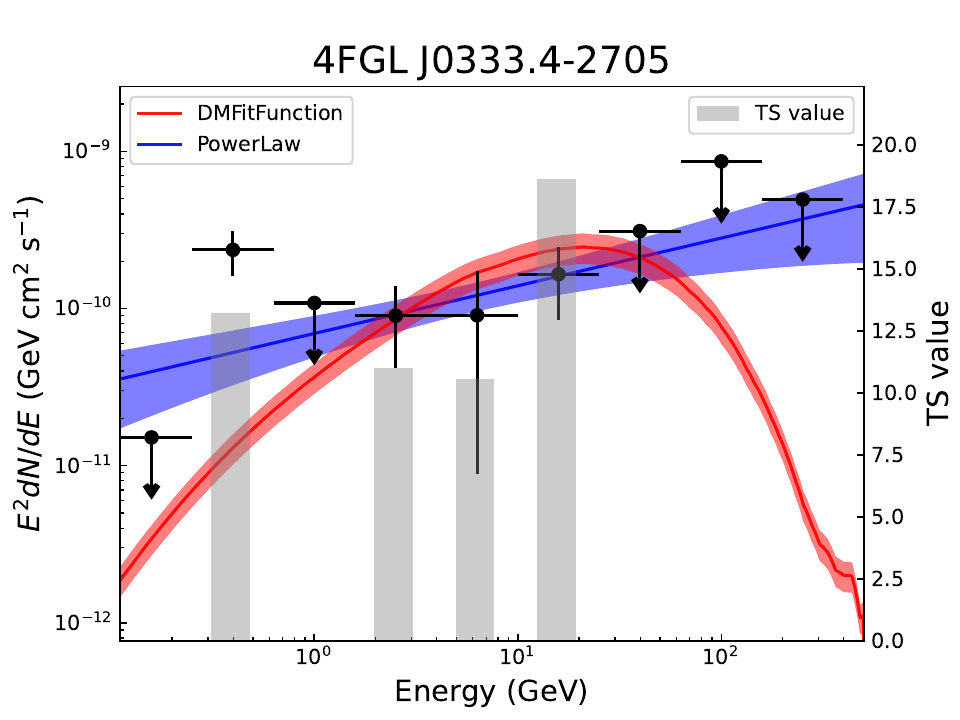}\\
\includegraphics[width=0.2\textwidth]{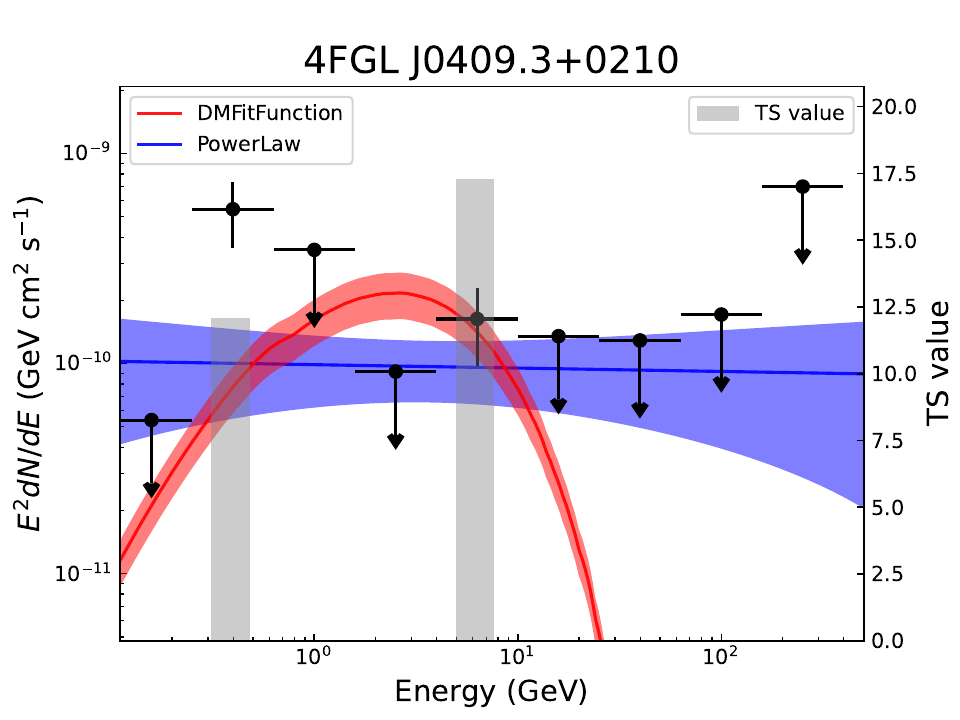}
\includegraphics[width=0.2\textwidth]{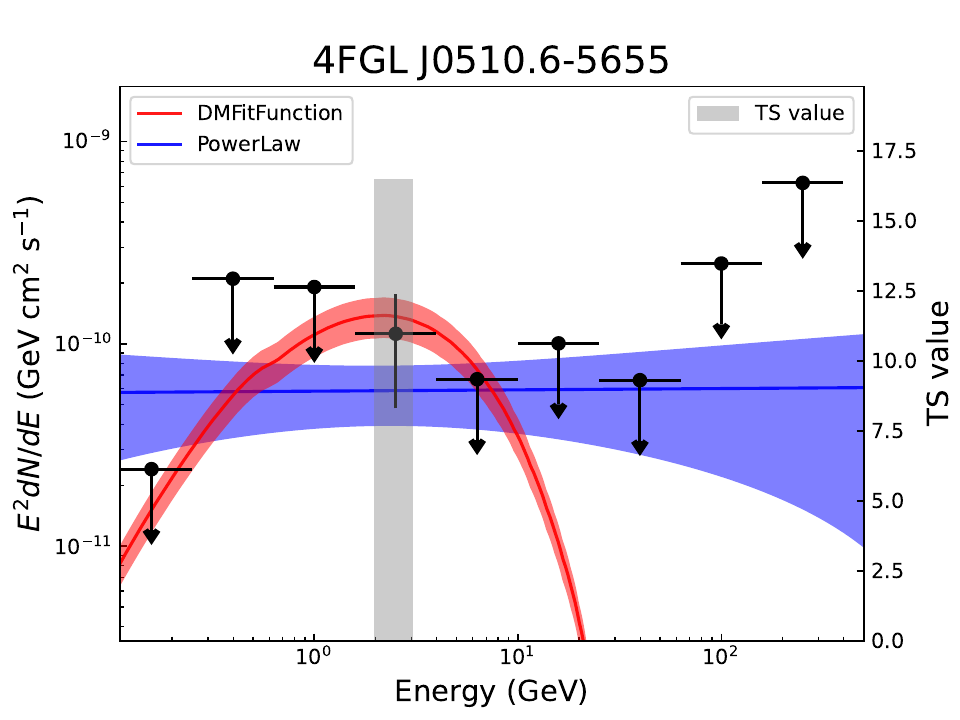}
\includegraphics[width=0.2\textwidth]{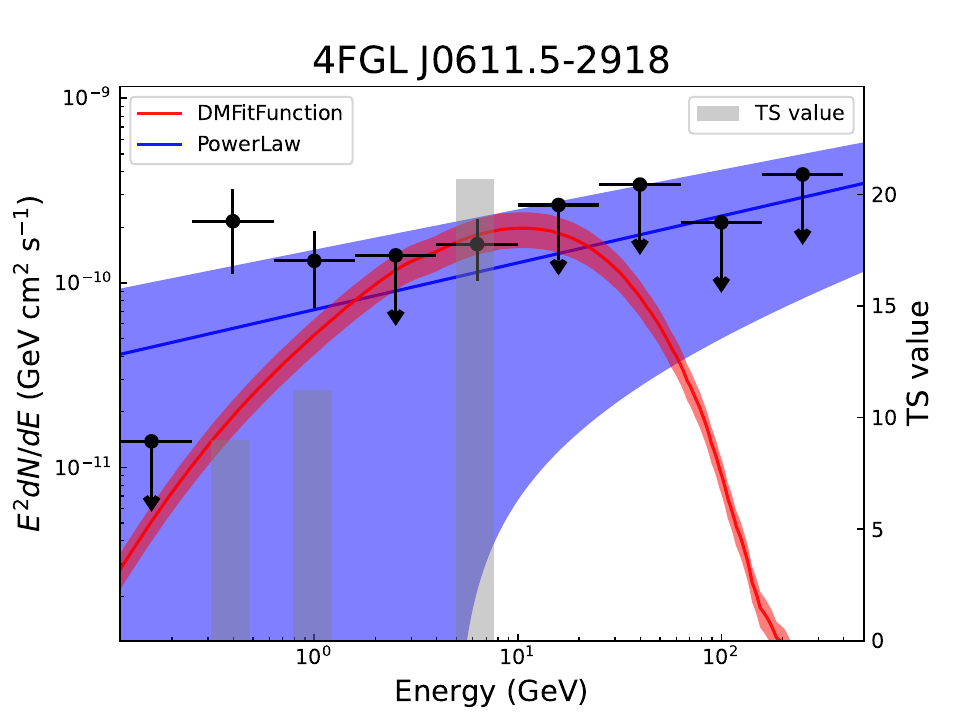}
\includegraphics[width=0.2\textwidth]{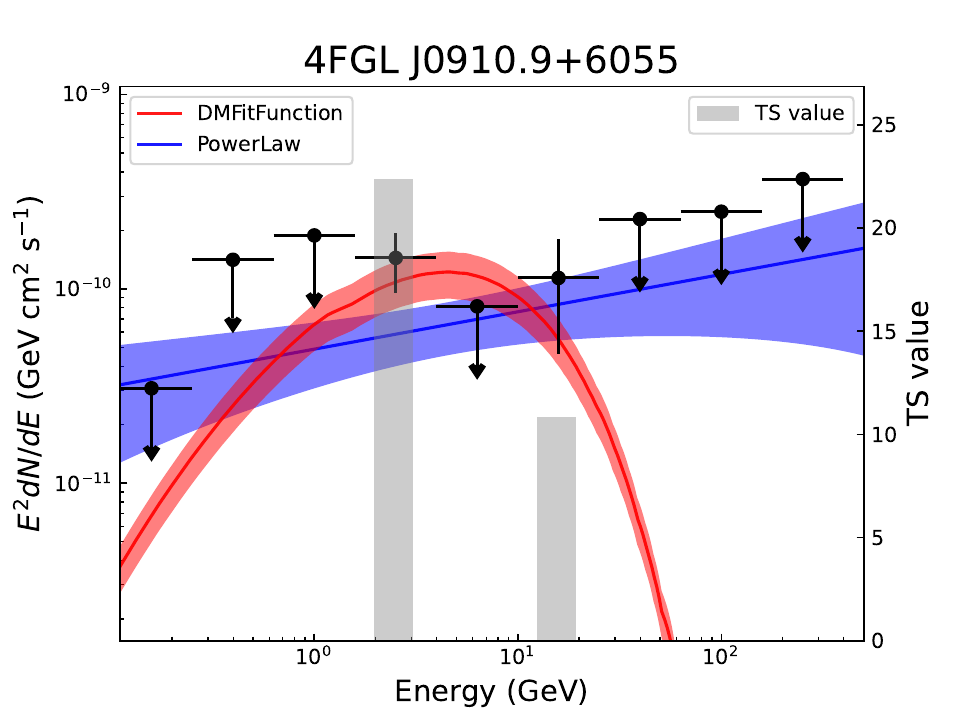}\\
\includegraphics[width=0.2\textwidth]{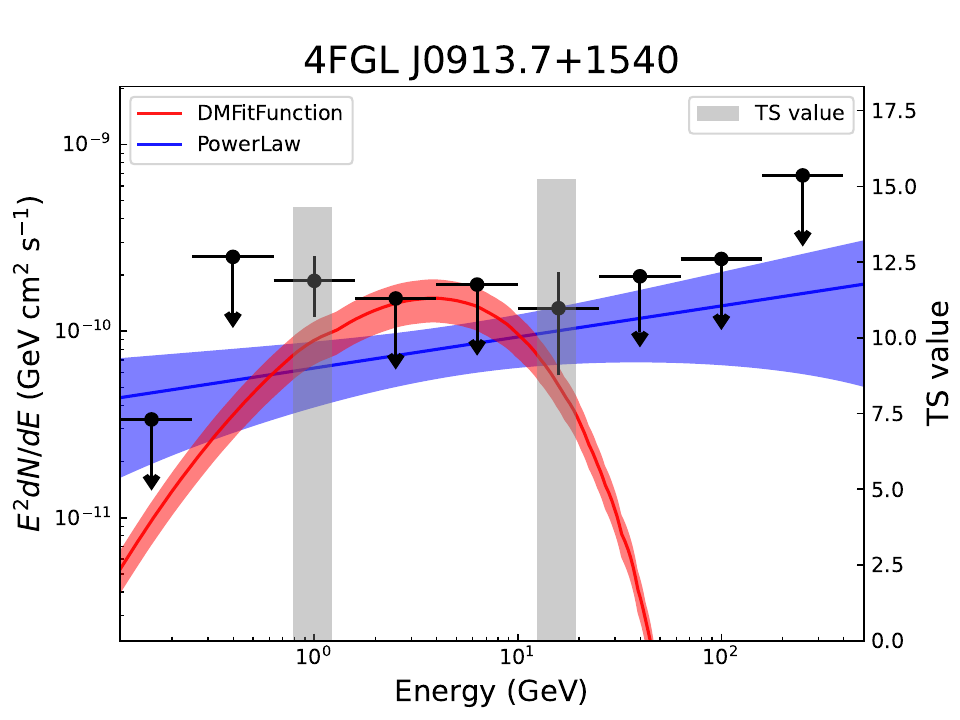}
\includegraphics[width=0.2\textwidth]{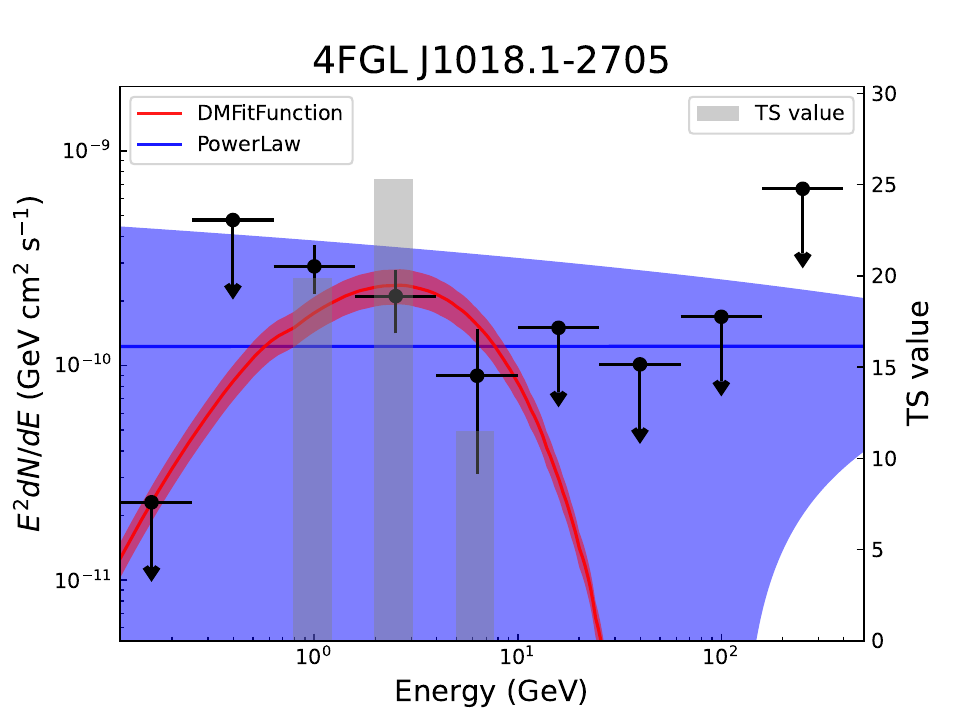}
\includegraphics[width=0.2\textwidth]{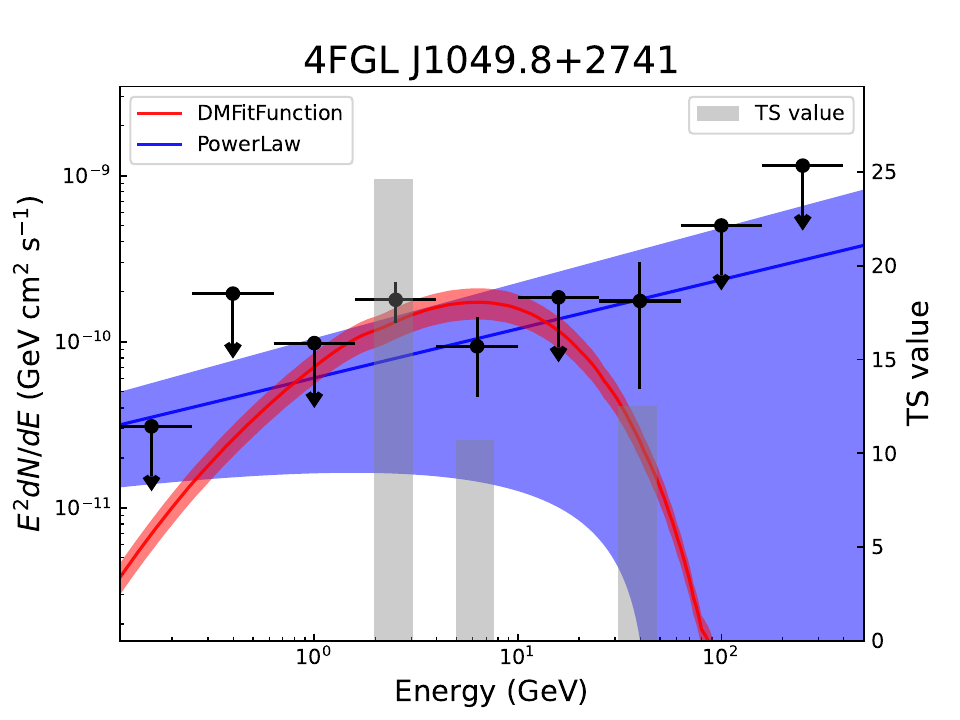}
\includegraphics[width=0.2\textwidth]{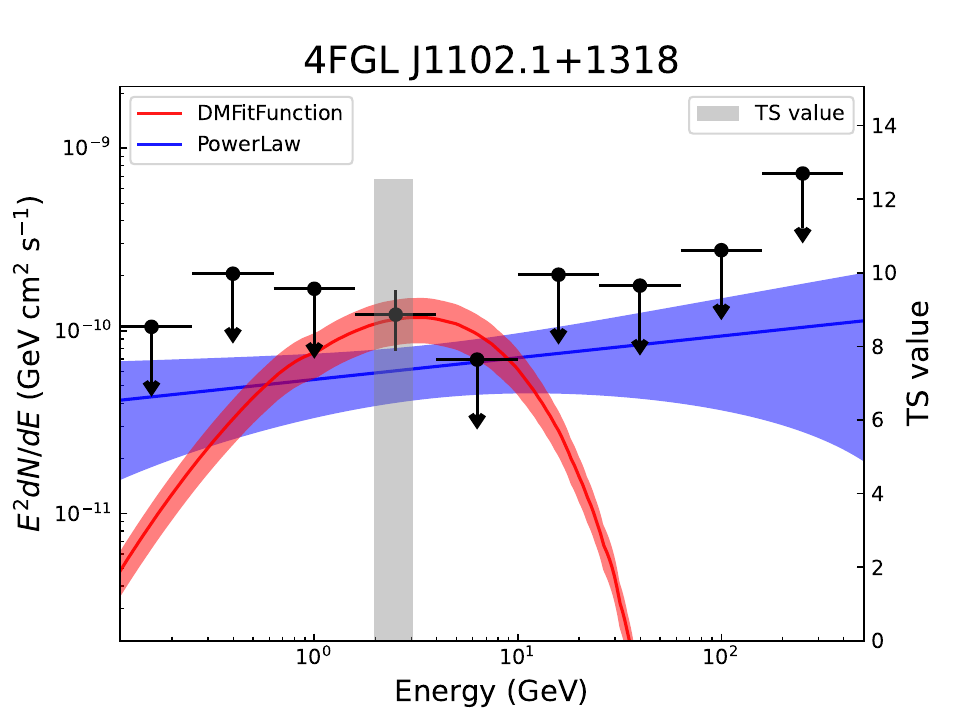}\\
\includegraphics[width=0.2\textwidth]{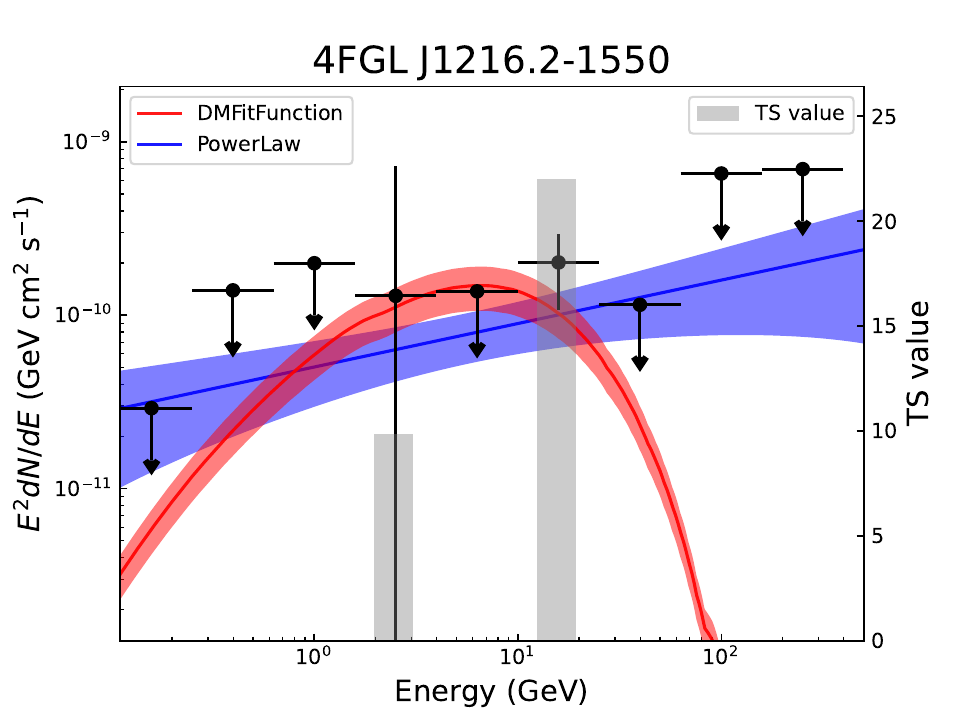}
\includegraphics[width=0.2\textwidth]{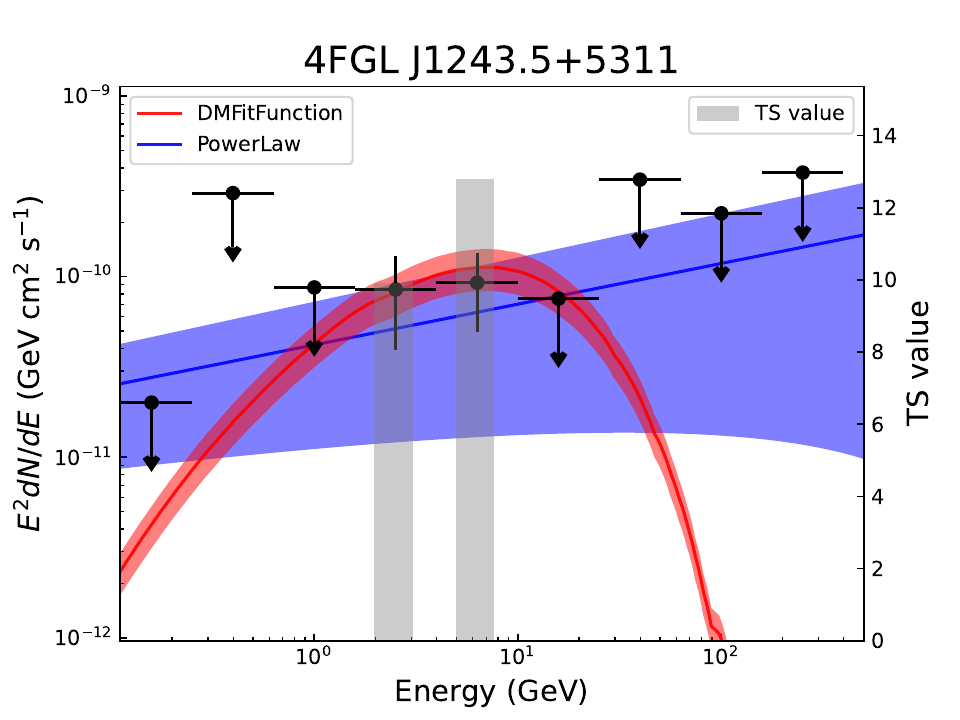}
\includegraphics[width=0.2\textwidth]{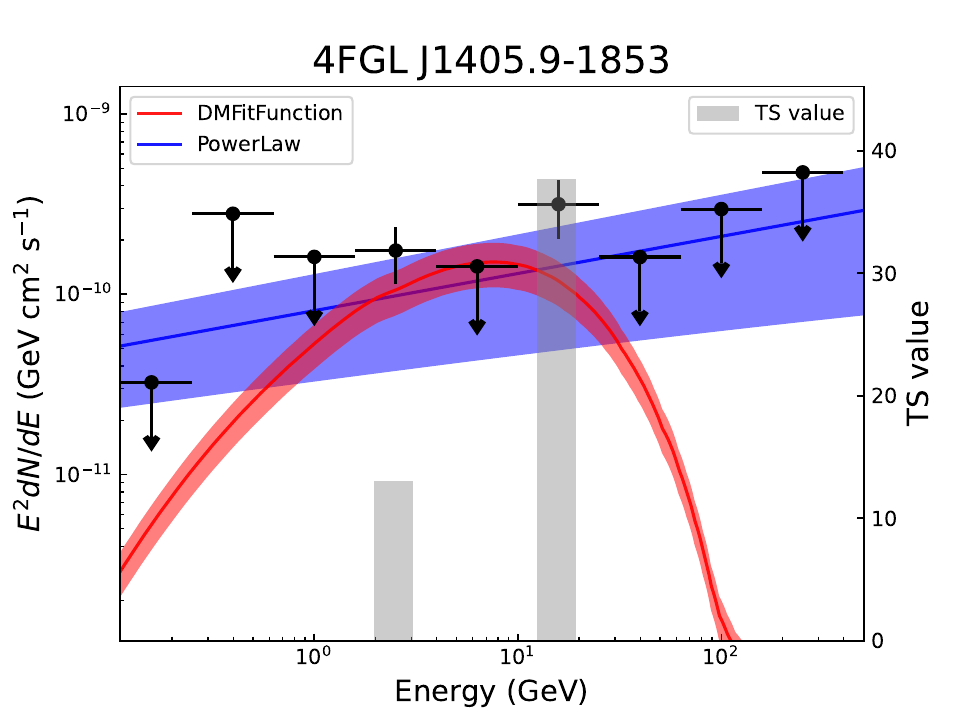}
\includegraphics[width=0.2\textwidth]{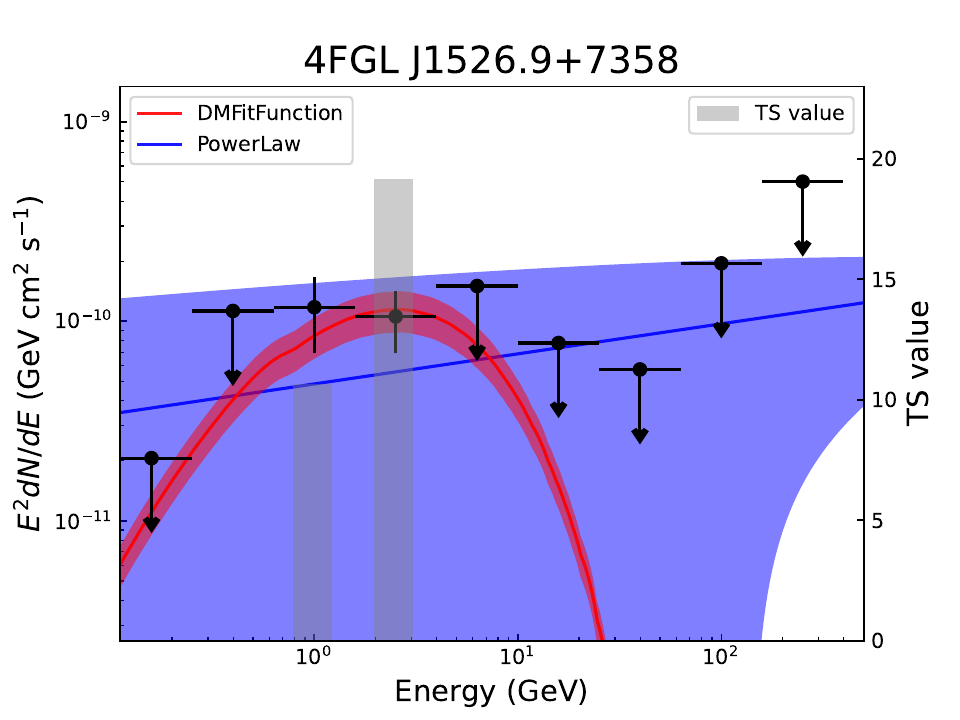}\\
\includegraphics[width=0.2\textwidth]{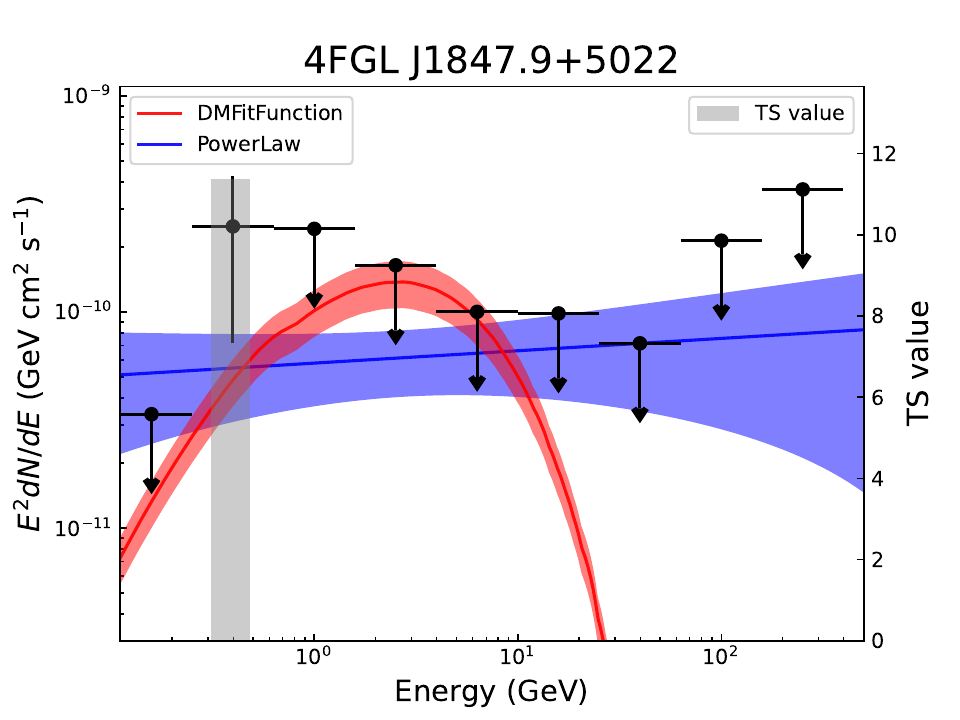}
\includegraphics[width=0.2\textwidth]{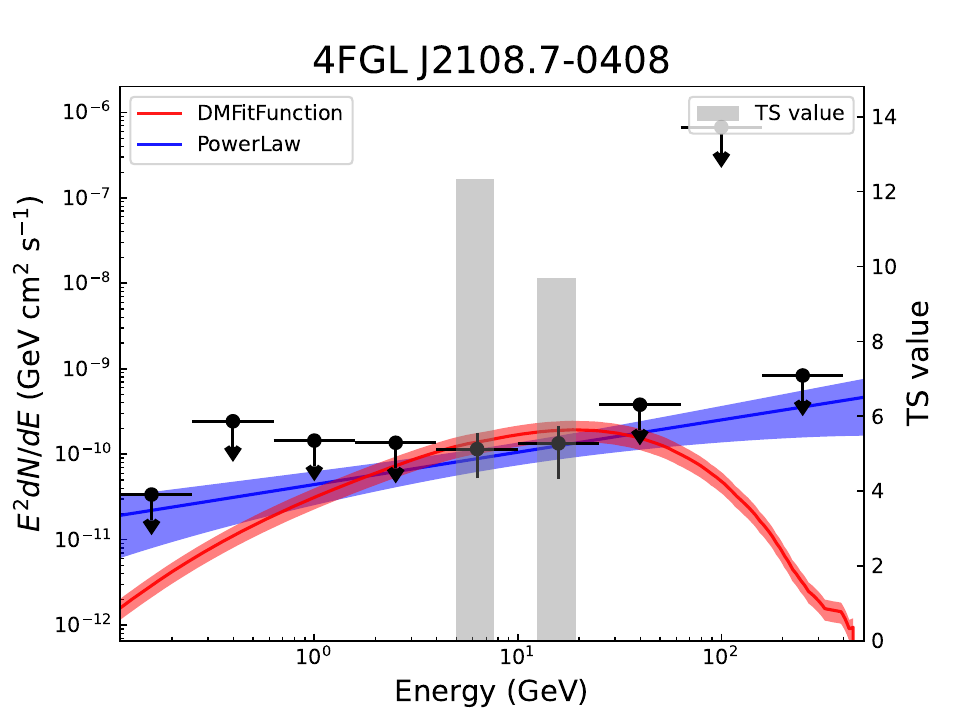}
\includegraphics[width=0.2\textwidth]{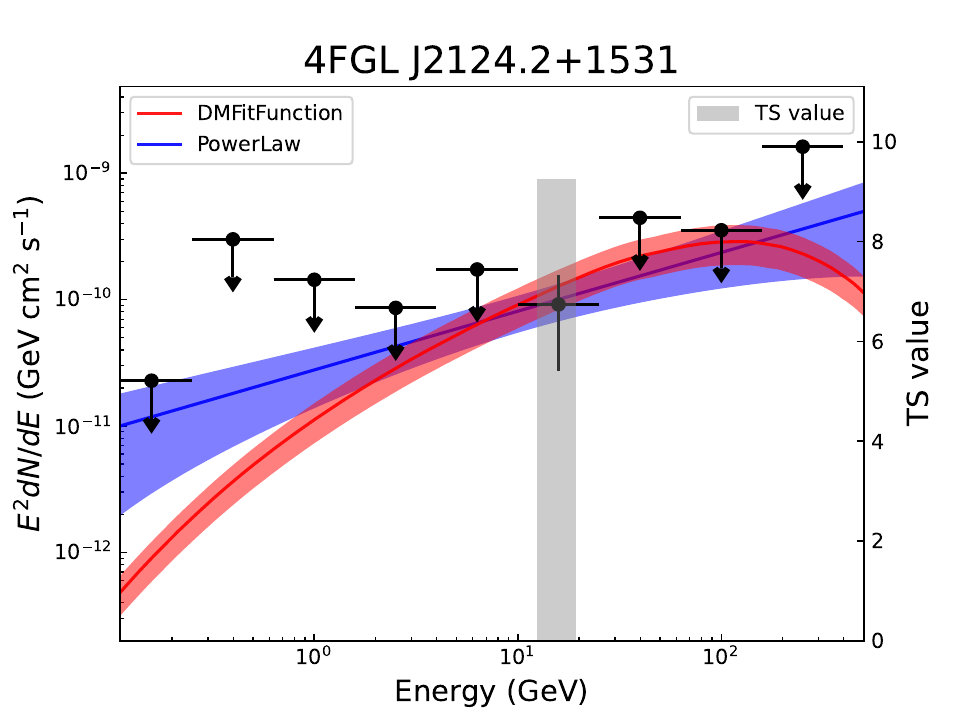}
\includegraphics[width=0.2\textwidth]{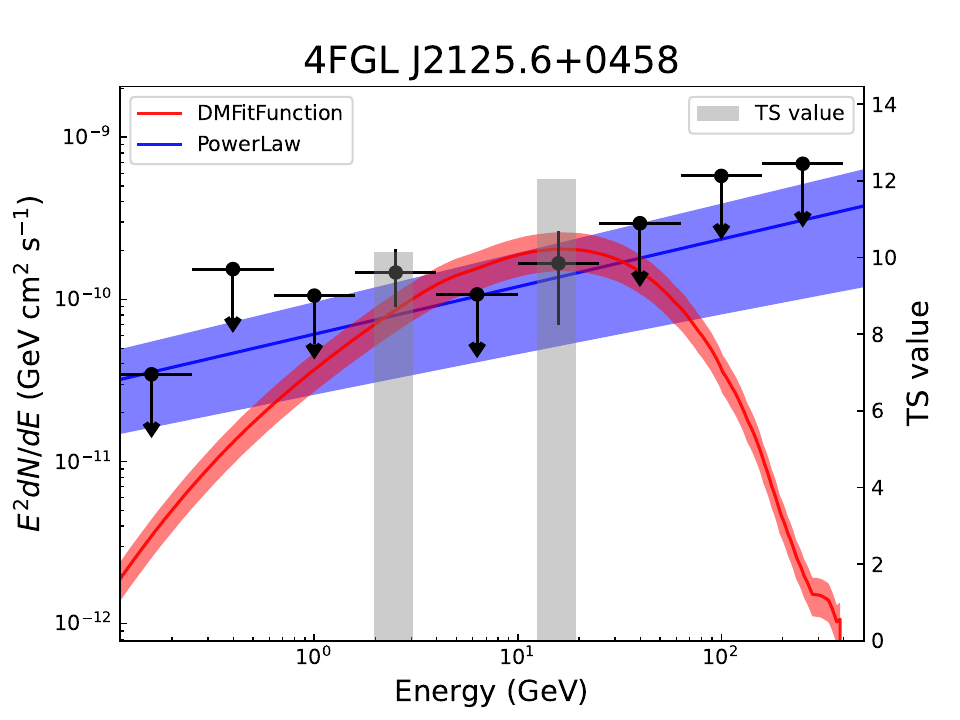}\\
\includegraphics[width=0.2\textwidth]{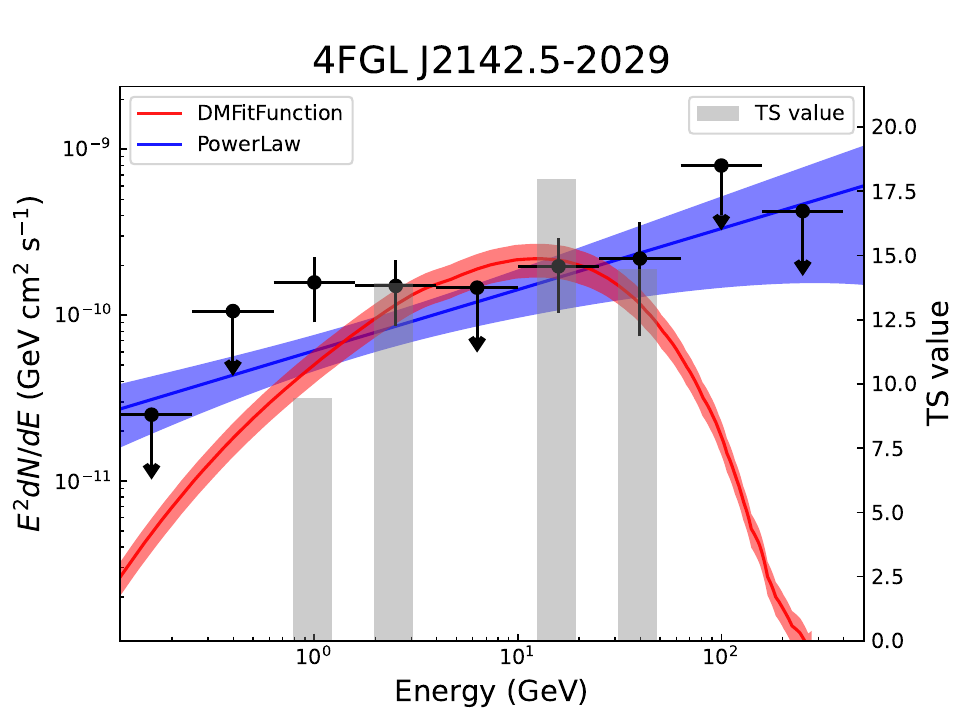}
\includegraphics[width=0.2\textwidth]{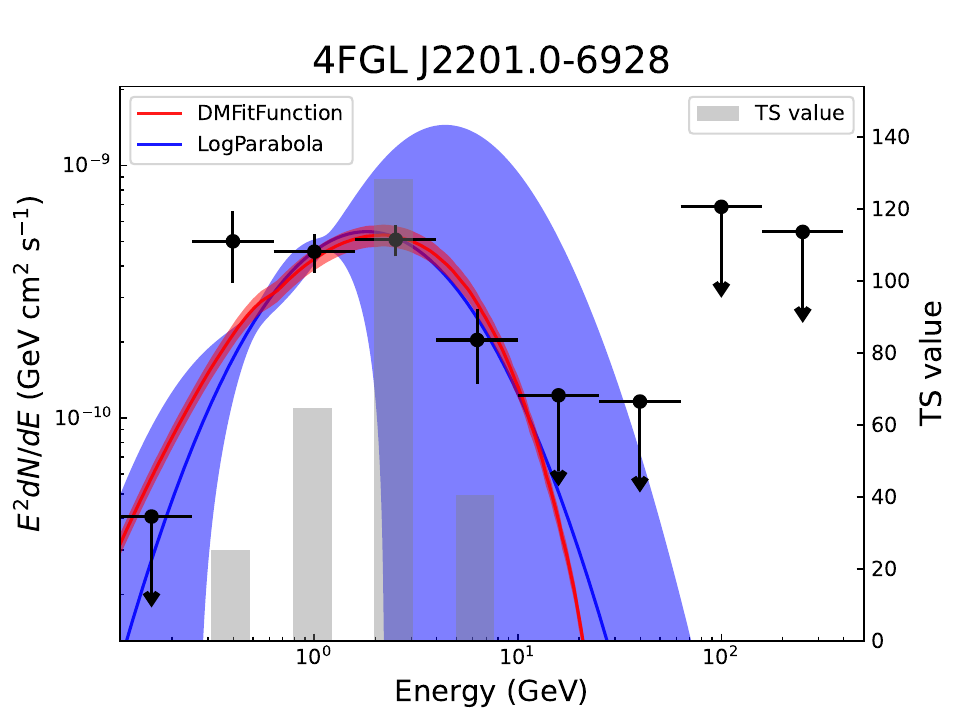}
\includegraphics[width=0.2\textwidth]{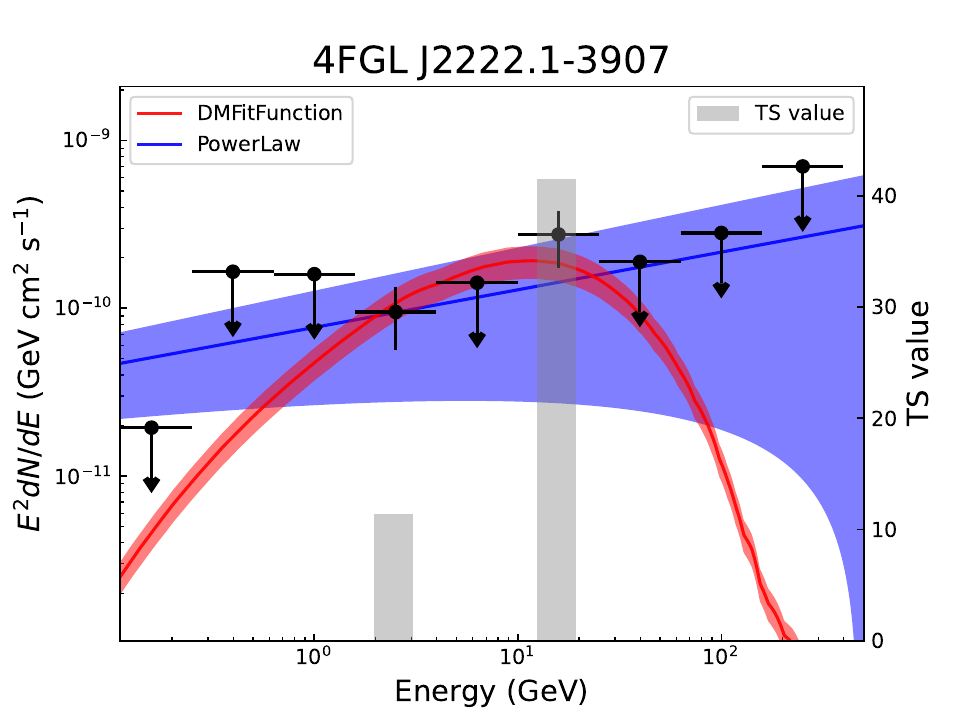}
\includegraphics[width=0.2\textwidth]{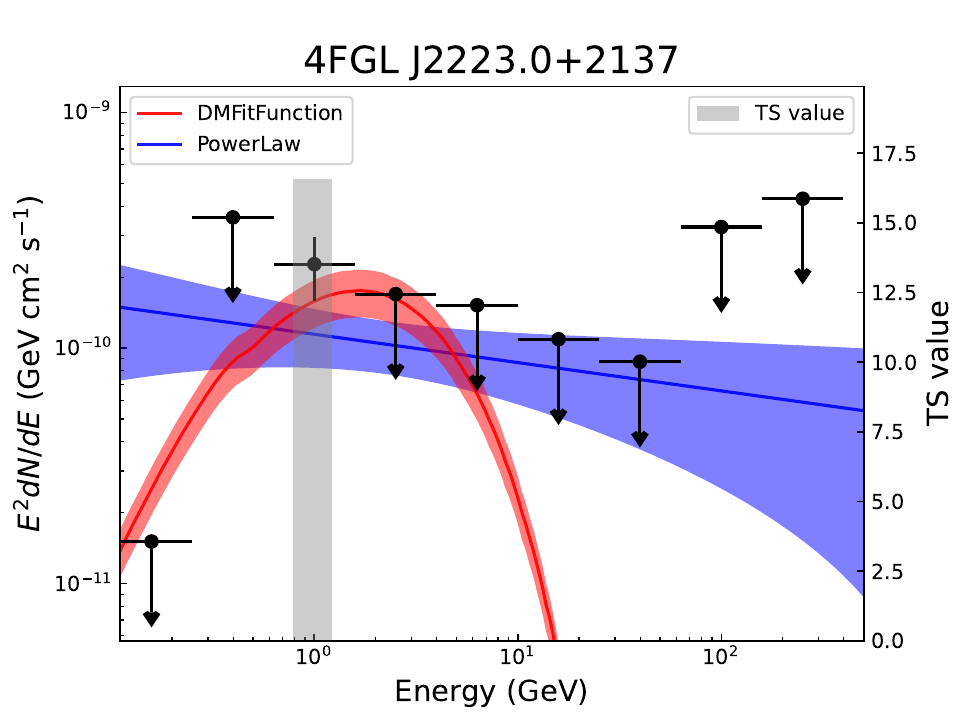}\\
\includegraphics[width=0.2\textwidth]{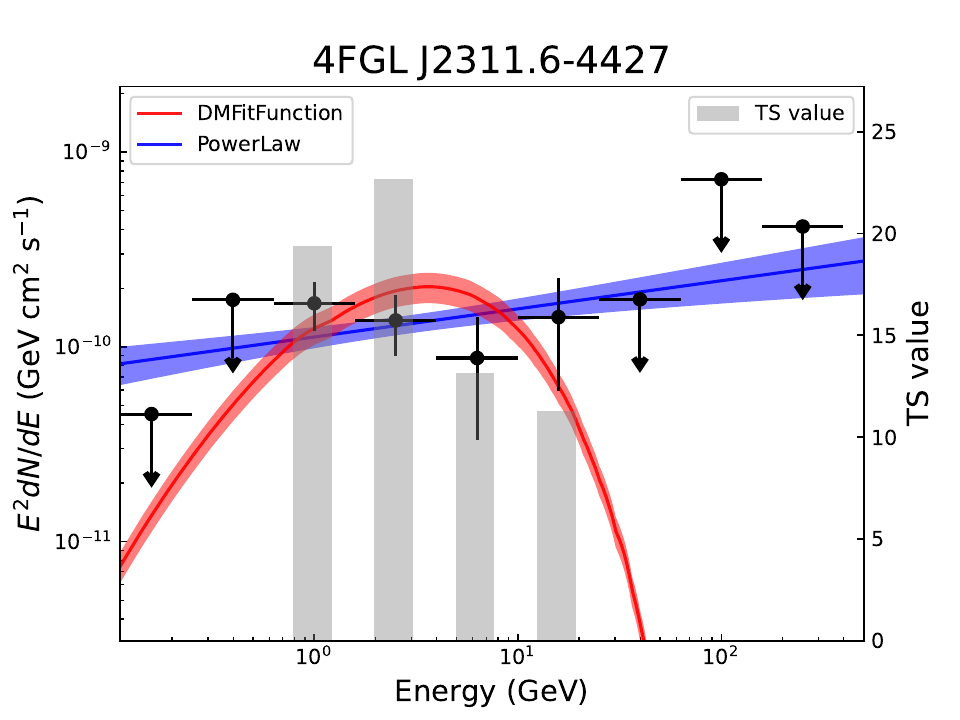}
\includegraphics[width=0.2\textwidth]{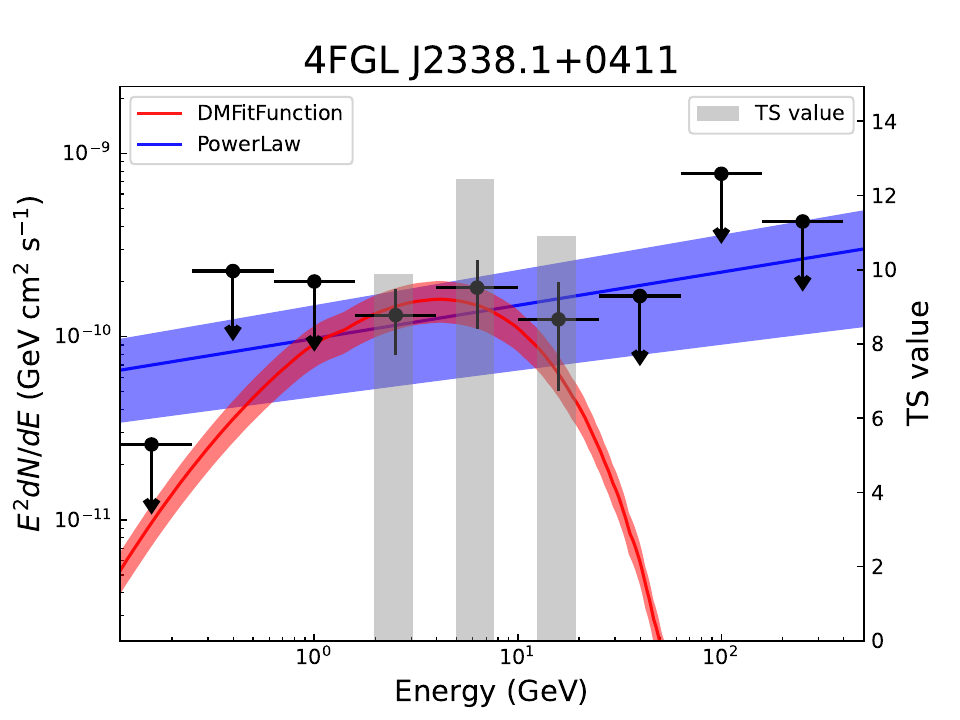}
\includegraphics[width=0.2\textwidth]{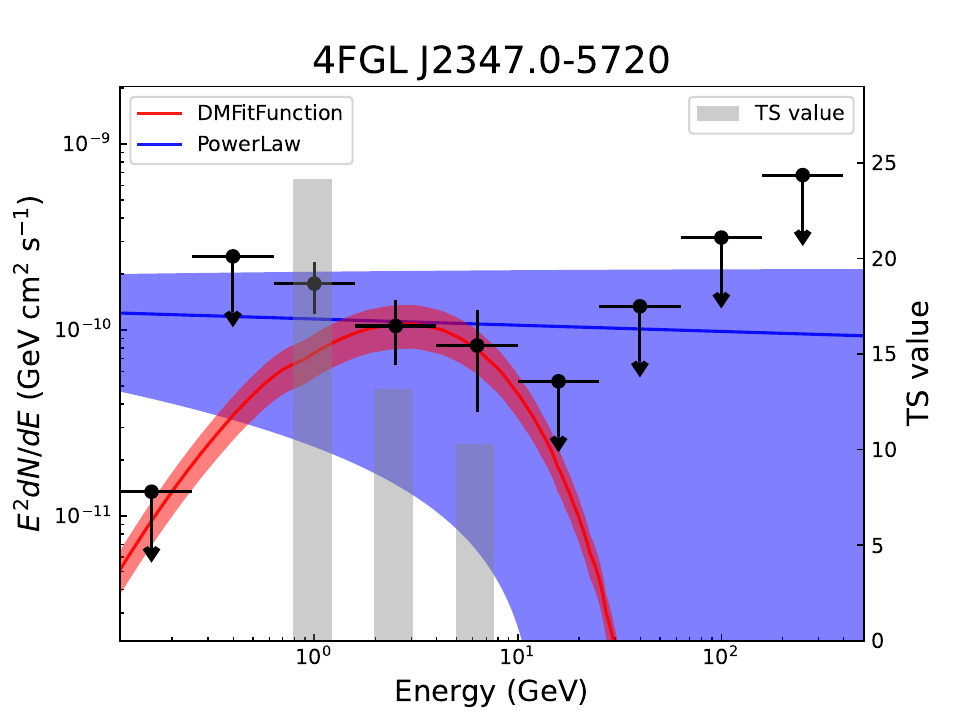}
\includegraphics[width=0.2\textwidth]{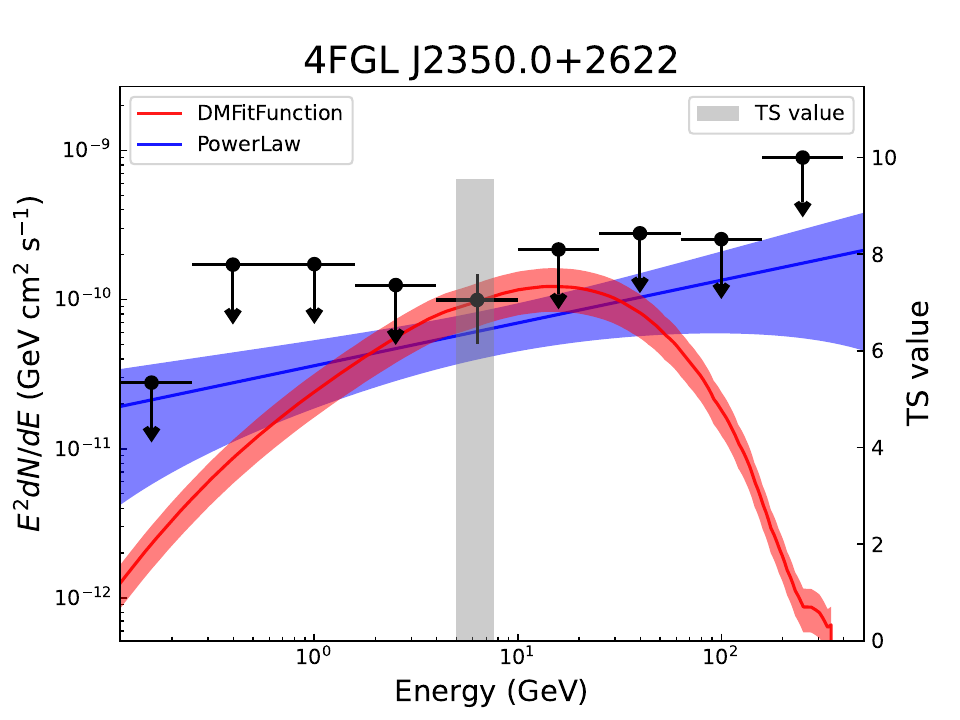}\\
\caption{The SEDs for all identified 32 DM subhalo candidates. The blue and red model lines correspond to the empirical function and the DM model respectively. The gray bars indicate the TS values in each energy bin and flux upper limits are estimated for bins with ${\rm TS} < 9$. }
\label{fig:seds}
\end{figure*}

We generate the spectral energy distribution (SEDs) of the candidates using a similar likelihood analysis procedure as described in Sec.~\ref{subsec:data_likelihood}. During the analysis, the target source model is replaced with a simple power-law function to estimate the flux in each of the 9 logarithmic energy bins. The resulting SEDs accompanied by the best-fit model lines are presented in Fig.~\ref{fig:seds}. The results show that the DM model aligns well with the SEDs. However, due to the limited statistics for these faint candidates, we cannot provide a further definitive conclusion at the current stage.

\section{Discussion}
\label{sec:dis}
An intriguing aspect of our results is that the 32 identified candidates appear to be entirely new and do not overlap with any previously reported candidates \cite{2015JCAP...12..035B,2019JCAP...11..045C,2020MNRAS.497.2486C}. For example, among the 24 candidates identified in Ref.~\cite{2015JCAP...12..035B} using the 3FGL catalog, 23 are still present in the 4FGL-DR4. Of these, 14 have since been associated (mostly with pulsars), while 9 remain unassociated. However, these 9 sources fail to meet our likelihood selection criteria and have been excluded. Several factors may explain this discrepancy. First, we use nearly twice as much observational data as the previous search, which could result in the DM annihilation model no longer adequately fitting the spectra of the sources due to the increased data accumulation. Second, the previous search relied on a different spectral fitting method, where SEDs were generated before being fitted by a DM model. The goodness of the fit is assessed by the $\chi^{2}$ value. This method is computationally efficient and practical in analyzing long observational period data, but the choice of energy bins would affect the fit results. In contrast, our approach uses an unbinned likelihood method, which avoids binning-related issues but is more computationally demanding for each target source. Additionally, our analysis considers both the DM model and an empirical function, applying a selection criterion of ${\rm ln}L_{\rm DM} > {\rm ln}L_{\rm emp}$. While this does not necessarily imply that the DM model is statistically superior to the empirical function, it at least suggests that the data is more consistent with the DM model.

\begin{figure*}
\centering
\includegraphics[width=1.0\columnwidth]{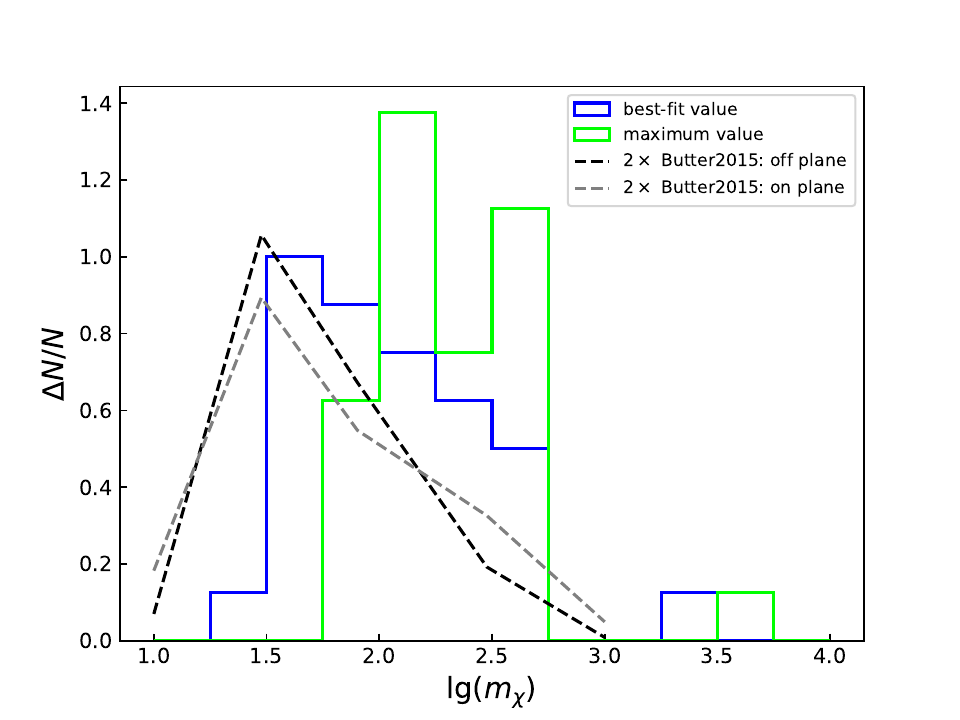}
\caption{The distribution of the DM particle mass for the 32 identified candidates. The blue line represents the best-fit DM masses from the likelihood spectral fitting, while the lime line corresponds to the maximum mass constraints derived from the fitting. For comparison, predictions from Ref.~\cite{2023JCAP...07..033B} are also included, adopting the most lenient threshold factor of $0.5$.}
\label{fig:mchi_distri}
\end{figure*}

Since the gamma-ray spectrum of the DM annihilation model depends strongly on the DM particle mass, once the annihilation channel is selected, it is insightful to examine the distribution of DM masses and compare it to predictions from Bayesian neural networks presented in Ref.\cite{2023JCAP...07..033B}. In our analysis, the reconstructed DM masses are reported as a best-fit value and a mass interval corresponding to a $2\sigma$ confidence level, as summarized in Tab.~\ref{tab:params}. It is worth noting that for some candidates, the derived minimum DM mass is $10\;{\rm GeV}$, which corresponds to the lower limit of the likelihood fitting range. This suggests that the minimum mass for these sources cannot be constrained by the available data. Hence, in constructing the DM mass distribution, we consider only the best-fit and maximum mass values. The resulting distributions are presented in Fig.~\ref{fig:mchi_distri}, alongside the predictions from Ref.\cite{2023JCAP...07..033B}. The comparison reveals that the distribution of best-fit DM masses follows a similar overall trend to the predictions, with the number of candidates decreasing as the DM mass increases. However, the decline in our distribution appears less steep, and no break is observed at $300\;{\rm GeV}$. Meanwhile, the maximum mass distribution significantly differs from both the best-fit mass distribution and the predictions, with the masses concentrated in the range of $[100, 500]\;{\rm GeV}$ and showing no clear trend. Several factors may explain the discrepancies between our results and the predictions. First, the relatively small sample size of 32 candidates introduces uncertainties related to the choice of binning. Second, our sample might not represent the complete set of candidates in the 4FGL-DR4 catalog, as we restricted our analysis to the $b \bar{b}$ annihilation channel, and other channels are not considered. 

The decision to focus on one annihilation channel was for a practical reason. As suggested in PPP4DMID \cite{2011JCAP...03..051C}, the spectral shapes of the hadronic channels $b \bar{b} / c \bar{c} / t \bar{t}$ are very similar, but slightly differ from those of the $e^{+} e^{-} / \mu^{+} \mu^{-}$ channels and the $\tau^{+} \tau^{-}$ channel. The differences in the predicted spectra across channels are not significant enough to be distinguished in the likelihood fitting, particularly for the faint sources that dominate the unassociated 4FGL-DR4 catalog. We tried to free the annihilation channel parameter of the DM model in the likelihood fitting, but the fits failed to converge due to the limited statistics. Further searches that incorporate other annihilation channels are likely to expand the candidate sample and provide a more comprehensive understanding of the potential DM subhalos population.

One caveat of our analysis is that we cannot rule out the possibility of the identified candidates actually being gamma-ray pulsars. This issue arises because the DM annihilation model exhibits a degeneracy with the exponential cutoff power-law (ECPL) model in explaining the candidate's observational spectra. The ECPL model is frequently used to describe the gamma-ray spectra of pulsars \cite{2012ApJ...747..121A}. Discriminating between these two models based solely on the SED is inherently challenging. Previous efforts to address this degeneracy have relied on indicators such as spectral curvature, peak energy, and detection significance \cite{2019JCAP...11..045C}. However, the most robust and effective approach to resolving this ambiguity is through multi-wavelength observations, including radio, optical, and X-ray bands. If a candidate is detected in other wavebands, it is less likely to be associated with DM. Future observations from advanced telescopes, such as the FAST \cite{2018IMMag..19..112L} and the Einstein Probe (EP, \cite{2015arXiv150607735Y}), will be crucial in investigating the true nature of these candidates.

\section{Acknowledgments}
We thank Yun-Feng Liang for the helpful suggestions and discussions. We acknowledge the data and software provided by the Fermi Science Support Center. This work is supported by the National Natural Science Foundation of China (No. 12403056).

\section{Data availability}
The Fermi-LAT data and software underlying this article are publicly available in the Fermi Science Support Center, at \url{https://fermi.gsfc.nasa.gov/ssc/data/access/} (data) and \url{https://github.com/fermi-lat/Fermitools-conda} (software).

\begin{table*}
\centering
\renewcommand{\arraystretch}{1.4}
\begin{tabular}{ccccccccc}
\hline
\hline
Source Name & $l$ & $b$ & $\int_{1\;{\rm GeV}}^{100\;{\rm GeV}} f {\rm d} E_{\rm \gamma}$ & $\mathcal{J}$ & $m_{\chi}$ & ${\rm TS}_{\rm DM}$ & ${\rm TS}_{\rm emp}$ & $\Delta \ln L$\footnote{$\Delta \ln L = \ln L_{\rm DM} - \ln L_{\rm emp}$} \\
(4FGL) & (deg) & (deg) & ($10^{-10}\;{\rm ph \,cm^{-2}\,s^{-1}}$) & ($10^{20}\;{\rm GeV^{2}\,cm^{-5}}$) & (${\rm GeV}$) &  &  & \\
\hline
J0024.1+2402 & $114.97$ & $-38.41$ & $2.16$ & $1.59_{1.34}^{1.84}$ & $134.20_{52.70}^{215.70}$ & $114.14$ & $100.94$ & $6.71$ \\
J0026.1-0732 & $104.76$ & $-69.52$ & $1.79$ & $3.81_{3.15}^{4.48}$ & $262.70_{187.17}^{338.23}$ & $105.42$ & $89.80$ & $38.95$  \\
J0046.9+0705 & $120.96$ & $-55.76$ & $1.22$ & $0.32_{0.24}^{0.40}$ & $52.56_{10.00}^{114.36}$ & $27.87$ & $16.37$ & $5.69$  \\
J0050.8+3330 & $122.79$ & $-29.36$ & $1.20$ & $0.39_{0.26}^{0.52}$ & $83.96_{10.00}^{175.71}$ & $27.96$ & $16.38$ & $5.80$  \\
J0102.0-6240 & $300.84$ & $-54.40$ & $1.25$ & $2.63_{2.01}^{3.24}$ & $309.06_{238.45}^{379.66}$ & $62.23$ & $55.66$ & $3.27$  \\
J0215.0-5330 & $278.26$ & $-59.44$ & $0.81$ & $0.56_{0.43}^{0.68}$ & $91.66_{10.00}^{180.18}$ & $41.78$ & $27.08$ & $10.73$  \\
J0225.5-5530 & $278.46$ & $-56.93$ & $2.57$ & $1.65_{1.38}^{1.93}$ & $110.57_{53.42}^{167.72}$ & $106.57$ & $109.70$ & $12.42$  \\
J0333.4-2705 & $222.41$ & $-54.10$ & $1.14$ & $5.83_{4.56}^{7.11}$ & $503.11_{456.55}^{549.67}$ & $70.66$ & $68.55$ & $1.17$  \\
J0409.3+0210 & $189.45$ & $-34.10$ & $1.78$ & $0.43_{0.33}^{0.54}$ & $47.80_{10.00}^{114.60}$ & $27.37$ & $17.98$ & $4.65$  \\
J0510.6-5655 & $265.27$ & $-36.07$ & $0.88$ & $0.22_{0.17}^{0.28}$ & $39.78_{10.00}^{80.16}$ & $30.98$ & $13.73$ & $8.54$  \\
J0611.5-2918 & $236.06$ & $-20.94$ & $1.17$ & $2.13_{1.66}^{2.60}$ & $234.77_{116.62}^{352.92}$ & $57.13$ & $48.97$ & $4.11$  \\
J0910.9+6055 & $154.41$ & $40.23$ & $1.05$ & $0.49_{0.36}^{0.63}$ & $92.44_{81.15}^{103.73}$ & $27.75$ & $20.69$ & $3.51$  \\
J0913.7+1540 & $214.02$ & $38.44$ & $0.77$ & $0.50_{0.37}^{0.63}$ & $77.29_{10.00}^{176.53}$ & $33.10$ & $28.35$ & $2.00$  \\
J1018.1-2705 & $265.61$ & $24.46$ & $1.95$ & $0.47_{0.39}^{0.56}$ & $47.96_{10.00}^{87.65}$ & $49.42$ & $31.56$ & $8.92$  \\
J1049.8+2741 & $204.65$ & $63.07$ & $1.21$ & $1.02_{0.80}^{1.24}$ & $132.44_{10.00}^{261.64}$ & $51.32$ & $49.74$ & $0.75$  \\
J1102.1+1318 & $235.91$ & $60.99$ & $0.87$ & $0.32_{0.23}^{0.41}$ & $63.02_{10.00}^{150.61}$ & $25.21$ & $20.43$ & $2.38$  \\
J1216.2-1550 & $290.67$ & $46.20$ & $1.04$ & $0.91_{0.65}^{1.17}$ & $137.12_{10.00}^{304.46}$ & $29.93$ & $22.75$ & $3.57$  \\
J1243.5+5311 & $125.63$ & $63.89$ & $0.73$ & $0.75_{0.55}^{0.94}$ & $148.07_{10.00}^{316.33}$ & $33.25$ & $26.57$ & $3.32$  \\
J1405.9-1853 & $326.36$ & $40.56$ & $1.36$ & $1.10_{0.79}^{1.40}$ & $161.73_{13.52}^{309.93}$ & $48.26$ & $42.91$ & $2.70$  \\
J1526.9+7358 & $110.05$ & $39.01$ & $0.83$ & $0.23_{0.18}^{0.29}$ & $48.71_{10.00}^{99.39}$ & $36.64$ & $22.25$ & $7.13$  \\
J1847.9+5022 & $79.83$ & $21.09$ & $1.01$ & $0.29_{0.22}^{0.35}$ & $49.24_{10.00}^{114.20}$ & $26.06$ & $14.78$ & $5.59$  \\
J2108.7-0408 & $46.07$ & $-32.24$ & $1.37$ & $4.02_{2.91}^{5.13}$ & $444.90_{414.20}^{475.61}$ & $41.48$ & $36.83$ & $2.32$  \\
J2124.2+1531 & $66.89$ & $-24.08$ & $0.88$ & $45.22_{29.45}^{60.98}$ & $3108.44_{2627.46}^{3589.42}$ & $30.54$ & $28.87$ & $0.85$ \\
J2125.6+0458 & $57.75$ & $-30.86$ & $0.92$ & $3.68_{2.69}^{4.67}$ & $385.89_{341.10}^{430.68}$ & $33.89$ & $30.87$ & $1.56$  \\
J2142.5-2029\footnote{Associated with 3FGL J2142.6-2029.} & $31.14$ & $-46.54$ & $1.45$ & $2.83_{2.18}^{3.48}$ & $279.89_{192.96}^{366.83}$ & $58.32$ & $57.85$ & $0.24$  \\
J2201.0-6928\footnote{Associated with 3FGL J2200.0-6930.} & $321.28$ & $-41.05$ & $4.57$ & $0.86_{0.78}^{0.95}$ & $39.64_{21.00}^{58.27}$ & $261.93$ & $262.28$ & $0.39$ \\
J2222.1-3907 & $2.13$ & $-56.80$ & $1.20$ & $2.27_{1.77}^{2.76}$ & $256.42_{235.94}^{276.90}$ & $72.73$ & $67.89$ & $2.55$ \\
J2223.0+2137 & $82.83$ & $-29.46$ & $1.12$ & $0.21_{0.16}^{0.26}$ & $30.02_{10.00}^{62.53}$ & $32.31$ & $25.65$ & $3.52$  \\
J2311.6-4427 & $345.45$ & $-63.56$ & $1.35$ & $0.63_{0.52}^{0.75}$ & $72.50_{14.70}^{130.31}$ & $72.00$ & $67.68$ & $2.36$  \\
J2338.1+0411 & $90.67$ & $-54.03$ & $1.52$ & $0.58_{0.43}^{0.73}$ & $84.18_{10.00}^{168.35}$ & $39.18$ & $38.10$ & $0.66$  \\
J2347.0-5720 & $319.21$ & $-57.71$ & $0.80$ & $0.25_{0.18}^{0.31}$ & $54.18_{10.00}^{117.63}$ & $26.24$ & $42.61$ & $358.20$  \\
J2350.0+2622 & $106.22$ & $-34.50$ & $0.88$ & $1.99_{1.34}^{2.64}$ & $348.43_{272.75}^{424.11}$ & $26.60$ & $22.67$ & $1.97$  \\
\hline
\hline
\end{tabular}
    \caption{4FGL-DR4 sources with spectra consisting of DM annihilation from DM subhalos.}
    \label{tab:params}
\end{table*}

\end{document}